\documentclass{article}
\usepackage{amsmath, amssymb, mathrsfs, bbm, mciteplus, abstract, paralist, url, graphicx, epstopdf}
\usepackage[english]{babel}
\usepackage[textwidth=442pt,textheight=646pt,a4paper]{geometry}

\newcommand{\real}[1]{\ensuremath{\mathbb{R}^{#1}}}

\title{{ \LARGE How Problematic is the Near-Euclidean Spatial Geometry of the Large-Scale Universe?}}
\author{\normalsize M. Holman\thanks{{\itshape \normalsize E-mail :} {\ttfamily \normalsize mholman8@uwo.ca}}\\
{\itshape \normalsize Rotman Institute of Philosophy and Department of Physics and Astronomy}\\ {\normalsize \itshape Western University, London, ON, N6A 5B8, Canada}}
\date{}

\begin{document}
\maketitle 
\begin{abstract}

\noindent Modern observations based on general relativity indicate that the spatial geometry of the expanding, large-scale Universe is very nearly Euclidean.
This basic empirical fact is at the core of the so-called ``flatness problem'', which is widely perceived to be a major
outstanding problem of modern cosmology and as such forms one of the prime motivations behind inflationary models.
An inspection of the literature and some further critical reflection however quickly reveals that the typical formulation of
this putative problem is fraught with questionable arguments and misconceptions and that it is moreover imperative to 
distinguish between different varieties of problem.
It is shown that the observational fact that the large-scale Universe is so nearly flat is ultimately no more puzzling than similar ``anthropic coincidences'', 
such as the specific (orders of magnitude of the) values of the gravitational and electromagnetic coupling constants. In particular, there is no fine-tuning problem in connection to flatness of the kind usually argued for.
The arguments regarding flatness and particle horizons typically found in cosmological discourses in fact address a mere \emph{single} issue underlying the standard FLRW 
cosmologies, namely the \emph{extreme} improbability of these models with respect to any ``reasonable measure'' on the ``space of all spacetimes''.
This issue may be expressed in different ways and a phase space formulation, due to Penrose, is presented here.
A horizon problem only arises when additional assumptions - which are usually kept implicit and at any rate seem rather speculative - are made.
\end{abstract}

\section{Introduction}

\begin{quote}
Though there never were a circle or triangle in nature, the truths demonstrated by Euclid would for ever retain their 
certainty and evidence.\\
David Hume \cite{Hume}                                                 						
\end{quote}

\noindent It is a commonplace that Euclid's theory of geometry, as layed down in the monumental, 13-volume \emph{Elements} \cite{Euclid},
reigned supreme for more than two millennia. The associated status of the theory, well reflected in the above quote by
Hume, in fact pertained to both its apparent faithful characterization of the geometry of ``physical space'' and its
axiomatic structure as a model for deductive reasoning\footnote{With respect to this latter aspect of Euclidean geometry,
Hume's verdict essentially remains accurate (i.e., theorems of Euclidean geometry have certainly not lost their validity in mathematics),
but it is unlikely that this is all he intended to say; he certainly did not contemplate the possibility of non-Euclidean circles or triangles.
Even better known than Hume's quote in this regard are of course Kant's notorious statements on the (synthetic) a priori truth of Euclidean geometry.}.
With the advent of non-Euclidean geometries in the nineteenth century, the geometrical nature of ``physical space''
became a question wide open and as famously emphasized by Riemann in his 1854 inaugural lecture \cite{Riemann},
this question is ultimately to be decided by experiment.
Nevertheless, Riemann's programme only became in-principle feasible in the twentieth century, after Einstein -
building heavily on some of Riemann's geometrical ideas - revolutionized the classical notions of space, time and eventually
gravitation through his two theories of relativity \cite{Einstein1,*Einstein2}.
Indeed, according to Einstein's theory of general relativity, gravitation is viewed no longer as a force, but instead, 
as a manifestation of curved \emph{spacetime} geometry.
In order to address the geometrical nature of ``physical spacetime'', it thus seems imperative to determine its
curvature. As is well known, non-Euclidean geometries first appeared as alternatives to Euclidean plane geometry
in which the parallel postulate was no longer assumed and it was found that they exist in two basic varieties,
depending on the sign of a single curvature constant, $K$, namely the \emph{elliptic plane} ($K > 0$),
characterized for instance by the convergent behaviour of initially parallel geodesics and the \emph{hyperbolic plane} ($K < 0$), 
characterized for instance by the divergent behaviour of initially parallel geodesics\footnote{In general, the (Gaussian) curvature of a 
two-dimensional surface of course need not be constant, although it locally still determines the geometry according to the three basic types characterized
by constant curvature. As will be seen however, there are good reasons to restrict attention to constant $K$ globally
and the entailed classification of geometries in the following.}.
Yet, the (Riemann) curvature of a four-dimensional spacetime geometry is in general specified in terms of twenty
independent, non-constant parameters and according to general relativity only half of these parameters are directly
(i.e., locally) expressable in terms of matter (or, more generally, non-gravitational ``source'') degrees of freedom
through Einstein's field equation.
In order to have a chance of arriving at an intelligible statement about spacetime's actual geometry, it is necessary, 
obviously, to resort to an approximation, based on what is (arguably) the big picture.
As will be recalled in section \ref{isotropy}, when this is done, general relativity quite remarkably \emph{predicts}
a characterization of the global geometry of spacetime in terms of a \emph{single} curvature constant, the meaning
of which is completely analogous to that in the intuitive, two-dimensional case.
Furthermore, and seemingly even more remarkable, the actually observed value of a particularly relevant cosmological
observable, $\Omega$, is in fact close to the value corresponding to a flat, i.e., Euclidean, geometry \cite{Adecs}.
This basic empirical fact is at the core of the so-called ``flatness problem'', which is widely perceived to be a major
problem of the standard Friedmann-Lema\^itre-Robertson-Walker (FLRW) cosmologies and as such forms one of the prime motivations behind the inflationary hypothesis
that is part of the current $\Lambda\mbox{CDM}$ ``cosmological concordance model''.
An inspection of the literature and some further critical reflection however quickly reveals that the typical formulation of
the flatness problem is fraught with questionable arguments and misconceptions and that it is moreover imperative to 
distinguish between different varieties of problem.
As will be shown in detail in section \ref{FP}, the observational fact that the large-scale Universe is so nearly flat 
is ultimately no more (or less) puzzling than similar ``anthropic coincidences'', such as the specific (orders of magnitude of the) values of the 
gravitational and electromagnetic coupling constants. In particular, there is no fine-tuning problem. The usual arguments
for the flatness and horizon problems found in inflationary discourses in fact address a mere \emph{single} 
issue underlying the standard FLRW cosmological models.
As will become clear in section \ref{IC}, this particular issue becomes a genuine theoretical problem, one that is argued
to require a dynamical mechanism such as inflation for its resolution, \emph{only} under additional assumptions, which are 
usually kept implicit and are at any rate speculative. It will also become clear however that if these assumptions 
are not made, this does not imply that the effectiveness of the FLRW approximation poses no problem.
In fact, a serious problem still remains, but it will require a type of solution that is radically different from
dynamical mechanisms such as inflation, as a result of its very nature.
For the benefit of the reader, section \ref{isotropy} consists of a brief, but mathematically concise review of
FLRW models and their empirical basis.
Geometrized units (i.e., $G_{\mbox{\scriptsize N}} := 1$, $c:=1$) are in effect throughout in what follows.

\section{Isotropic Cosmologies}\label{isotropy}

\noindent One of the most startling consequences of general relativity is that it almost \emph{inevitably}
leads to the notion of a dynamically evolving Universe. As is well known, it was Einstein himself who, when confronted
with this situation in an attempt to model the Universe's spatial geometry as a three-sphere, with matter distributed
uniformly, sought to avoid this particular implication of his theory by introducing an additional, ``cosmological constant'' 
term into the field equation \cite{Einstein3}. By carefully adjusting the value of the cosmological constant to a specific magnitude,
the picture of a static, non-evolving Universe could be retained, but only so at the expense of conflicting with ``naturalness''
- as soon became clear through independent works of Friedmann \cite{Friedmann} and Lema\^{\i}tre \cite{Lemaitre}, most notably - and eventually of course
also firm observational evidence in the form of Hubble's law \cite{Hubble}.
Although Einstein's particular interpretation of a cosmological constant had thus become falsified by the late 1920s,
whereas his arguments for a closed, elliptic spatial geometry were also generally regarded as unconvincing, what did
survive in essentially unmodified form from the effort, was the picture of a Universe with matter distributed uniformly.\\
That is to say, underlying the dynamically evolving solutions obtained by Friedmann and Lema\^{\i}tre in the 1920s
(or, for that matter, the current standard model of cosmology, which is fundamentally based upon these solutions), 
is the assumption that on sufficiently large spatial scales, the Universe is essentially \emph{isotropic}, i.e., essentially 
``looks the same'' in every direction at every point of ``space''.
Obviously, it is important to have a rough sense of the cosmological scales at which isotropy is thought to become effective.
For instance, while typical galactic scales are absolutely enormous compared to any human scale (taking our Milky Way
Galaxy - which measures $\sim$ 100,000 ly (roughly $10^{21}$ m) in diameter and is estimated to contain 
some $10^{11}$ stars, distributed in the form of a spiral-like, thin disk - as more or less representative, photons from 
stars on the opposite side of the Galaxy and incident upon Earth now, for instance, were emitted long before any human civilization 
existed), such scales are still tiny when compared to the overall scale of the observable Universe, which is 
another five orders of magnitude larger\footnote{Taking the Milky Way diameter as the unit of length, our
nearest galactic neighbour, Andromeda (M31 in the Messier classification), is at a distance of more than twenty length
units, whereas the most distant galaxy known to date, GN-z11 (the number representing the galaxy's redshift $z$), is
more than 300,000 galactic length units away.
It is easy to forget that the very question as to whether other galaxies even existed was settled only in 1925, when Hubble
was able to reliably determine distances to Cepheid variable stars in the Andromeda Nebula for the first time and thereby 
established the extra-Galactic nature of these variables.}.
While galaxies are also found to (super-)cluster on a wide range of scales and while at the same time there appear to
be large regions effectively devoid of any galaxies (and apparently matter), observations indicate that on the
largest cosmological scales, the Universe is isotropic to a very good degree of approximation.
In particular, empirical support for isotropy comes from (i) direct astronomical observations of remote galaxies within 
optical and radio regions of the electromagnetic spectrum (including measurements of Hubble redshifts), (ii) indirect 
observations, such as the distribution pattern of cosmic rays incident upon Earth, and, most significantly, (iii) the
\emph{extra-ordinary} uniformity - i.e., up to a few parts in $10^5$ - of the \emph{Cosmic Microwave Background} (CMB) \cite{Adecs2,*WuLaRe,*Longair}.\\
Although this observational evidence strictly speaking only indicates that the large-scale Universe is
isotropic with respect to \emph{us} (i.e., the Milky Way Galaxy), it is commonly considered bad philosophy 
since the time of Copernicus to take any such circumstance as evidence that we somehow find ourselves at a 
privileged (spatial) position.
This very reasonable position, which thus in essence states that the large-scale Universe is spatially \emph{homogeneous},
i.e., that at any ``instant of time'', the observable features of other regions of the Universe (measured at sufficiently 
large scales) are effectively identical to those of our region, is also well supported by experimental data \cite{Hoggcs,*Sarkarcs}.
Empirical evidence, together with common sense, thus strongly suggests that, to a very good degree of approximation, the large-scale 
observable Universe is spatially isotropic\footnote{The scale at which the transition to homogeneity occurs corresponds to $\sim 300 \,  \mbox{Mly}$ -
or about 3,000 ``galactic units'' \cite{Hoggcs,*Sarkarcs}. Since any two points on a hypersurface of
homogeneity are ``equivalent'', (spatial) isotropy with respect to any point on such a surface implies isotropy with 
respect to all points (see also note \ref{homogeneity}).
A slightly different justification for the ``Copernican Principle'' (i.e., spatial homogeneity) that is sometimes
given is based on the observation that spherically symmetric solutions in general relativity are symmetric either with 
respect to at most two points or with respect to \emph{every} point \cite{HawkingEllis} and the former would only be 
consistent with the observed isotropy in an essentially anthropocentric universe.
This way of formulating matters strongly suggests that, assuming general relativity to be correct, the large-scale
Universe is spatially isotropic - and, by implication, homogeneous - everywhere, i.e., also beyond the Hubble radius.
In addition to the more or less direct empirical evidence in favour of isotropy already mentioned,
more sophisticated tests of the Copernican principle (so far confirmed) have also been proposed over the years \cite{Uzancs,*Cliftoncs}.}.
As it turns out, such a uniformity condition is mathematically very restrictive and completely fixes the form of the
spacetime metric, $g_{ab}$. For future reference, the most characteristic features of a spatially 
isotropic spacetime $(M,g_{ab})$ are briefly recalled here\footnote{A spacetime, $(M,g_{ab})$, is said 
to be (spatially) isotropic if there exists a congruence of timelike curves, such that at each point $x \in M$, there 
exists a subgroup of isometries of $g_{ab}$, which leaves $x$ and vectors tangent to the congruence at $x$ fixed, but 
which acts transitively on the subspace of vectors orthogonal to the congruence at $x$ \cite{Wald}. In a similar fashion, $(M,g_{ab})$
is said to be (spatially) homogeneous if $M$ can be foliated by a one-parameter family of spacelike hypersurfaces, $\Sigma_{\tau}$,
such that for all $\tau$, any two points in $\Sigma_{\tau}$ can be connected by an isometry. Hence, all points in any given
hypersurface of homogeneity are ``equivalent''. Isotropy is a more stringent condition than homogeneity, since the former
implies the latter but not vice versa (intuitively, any inhomogeneities between large-scale regions would appear as
large-scale anisotropies to observers equidistant from these regions so that an isotropic spacetime must be homogeneous \cite{Rindler}).\label{homogeneity}}. 

\begin{figure}
\begin{center}
\includegraphics[width=80mm]{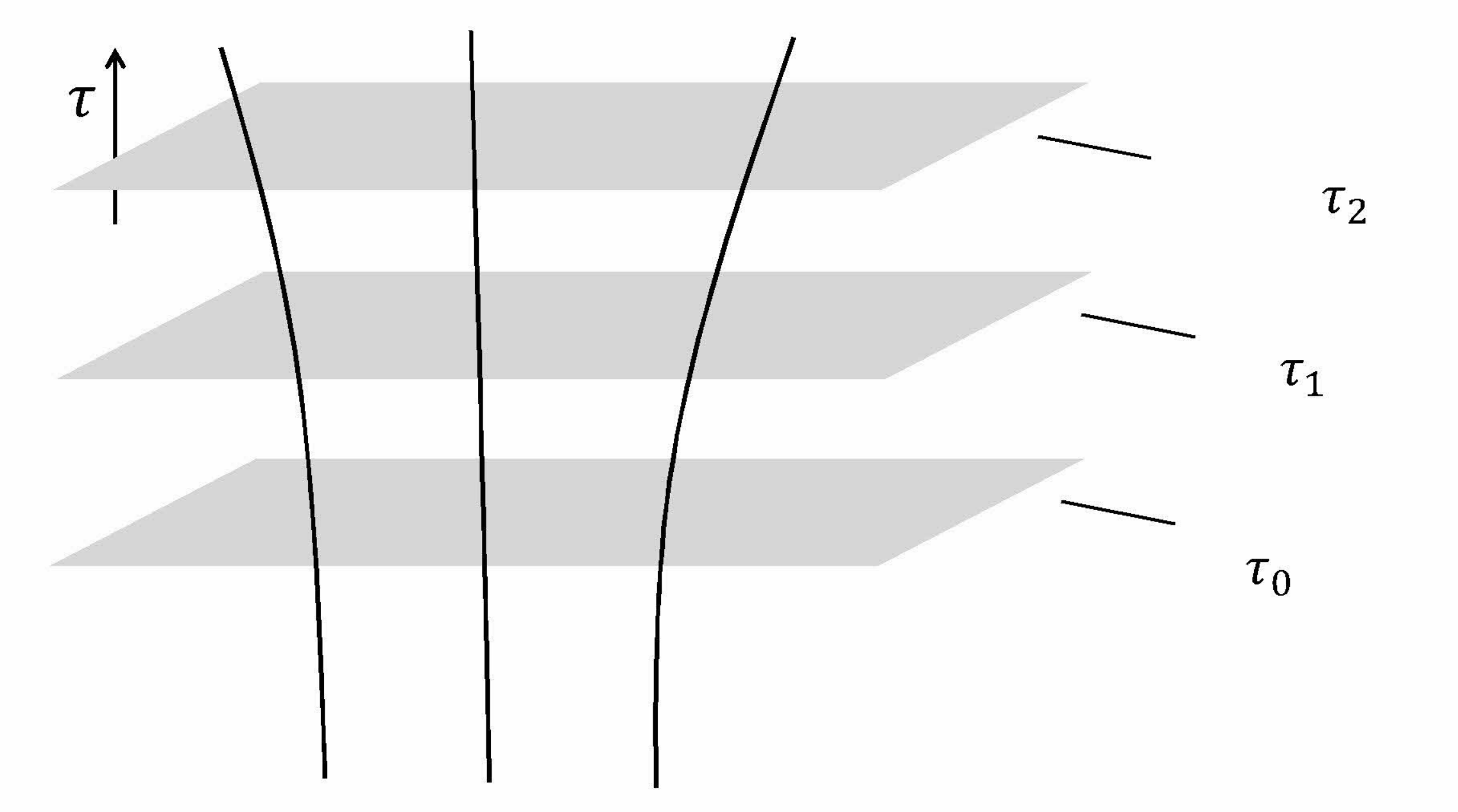}
\caption{Foliation of a FLRW spacetime by hypersurfaces of homogeneity, $\Sigma_{\tau}$, representing different ``instants
of time'', $\tau$. These spatial geometries are orthogonally intersected by a congruence of timelike geodesics, which are the
worldlines of isotropic observers (representing individual galaxies), and have constant curvature labeled by a 
discrete index $k \in \{ -1 , 0 , 1\}$.}\label{foliation}
\end{center}
\end{figure}

\begin{itemize}
\item[(i)] $M$ is foliated by a family of spacelike hypersurfaces, $\Sigma_{\tau}$ , representing different ``instants of 
time'', $\tau$, and there exists a preferred family of isotropic ``observers'' (i.e., galaxies) represented by a congruence
of timelike geodesics with tangents, $u^{a}$, orthogonal to the surfaces of simultaneity (cf. Fig. \ref{foliation} and footnote \ref{homogeneity}).
\item[(ii)] Each spatial hypersurface is a space of \emph{constant curvature} and the metric takes the \emph{Friedmann-Lema\^itre-Robertson-Walker (FLRW)} form\footnote{In addition to the
contributions of Friedmann and Lema\^{\i}tre already mentioned, Robertson \cite{Robertson} and Walker \cite{Walker} 
independently showed that the metric (\ref{RWmetric}) applies to all locally isotropic spacetimes.}
\begin{equation}\label{RWmetric}
ds^2 \; = \; - d \tau^2 \: + \: a^2(\tau) \left( \frac{dr^2}{1 - kr^2} \: + \: r^2 d \omega^2 \right)
\end{equation}
with $\tau$ denoting the proper time of galaxies, $a(\tau)$ representing an overall \emph{scale-factor} of 
the homogeneous three-geometry, $\Sigma_{\tau}$ and $k$ a curvature constant which equals either $-1$, $0$, or $+1$, depending
on whether the spatial geometry is respectively \emph{hyperbolic}, \emph{Euclidean}, or \emph{elliptic} ($d \omega^2 = d \theta^2 + \sin^2 \theta d \phi^2$,
with $(r,\theta,\phi)$ denoting spherical coordinates as usual). In other words, the spatial geometry is essentially
fixed once a value for $k$ is specified\footnote{As a consequence of their constant curvature, any two hypersurfaces
$\Sigma_{\tau}$, $\Sigma_{\tau'}$ with the same value of $k$ are locally isometric.
However, two foliations by hypersurfaces with the same value of $k$ could in principle still represent topologically distinct
spatial geometries (a flat, $k=0$ geometry could for instance either be ``open'', corresponding to standard infinite Euclidean three-space,
or ``closed'', corresponding to e.g.\ a three-dimensional torus). Although it is often argued, either explicitly
or implicitly, that examples of this kind demonstrate that cosmology faces a serious, if not irreconcilable, problem of underdetermination,
it seems that such arguments may underestimate human ingenuity. In addition, it is also not inconceivable that
a future theory of quantum gravity could decide spatial topology.\label{topology}}.
\item[(iii)] The metric (\ref{RWmetric}) is \emph{conformally flat}, i.e., can be expressed in the form $\Upsilon^2 \eta$, where $\eta$
denotes the standard Minkowski metric and $\Upsilon$ is a strictly positive (smooth) function.
Hence, the Weyl tensor vanishes identically, $C^{a}_{\mspace{10mu} bcd} = 0$, and there are no proper gravitational degrees of
freedom in a FLRW spacetime.
\item[(iv)] By virtue of the symmetry, the stress-energy tensor, $T_{ab}$, necessarily takes the \emph{perfect fluid form}
\begin{equation}\label{isofluid}
T_{ab} \; = \; \rho u_a u_b \: + \: P (g_{ab} + u_a u_b)
\end{equation}
with $\rho$ and $P$ respectively denoting the (spatially constant) energy density and pressure of the fluid, as a 
result of which the ten, a priori independent, components of Einstein's field equation
\begin{equation}\label{Einstein}
G_{ab} \: + \: \Lambda g_{ab} \; := \; R_{ab} \: - \: \frac{1}{2}R g_{ab} \: + \: \Lambda g_{ab} \; = \; 8 \pi T_{ab} 
\end{equation} 
(with $R_{ab}$, $R$ and $\Lambda$ respectively denoting, as usual, the Ricci tensor, scalar curvature and cosmological constant)
reduce to just two independent equations
\begin{eqnarray}
\frac{3\dot{a}^2}{a^2} & = & 8   \pi   \rho \: + \: \Lambda \: - \: \frac{3k}{a^2} \label{RWmetric1}  \\
\frac{3\ddot{a}}{a}    & = & - 4 \pi ( \rho \: + \: 3 P) \: + \: \Lambda           \label{RWmetric2}
\end{eqnarray}
where the dot denotes differentiation with respect to proper time. Eq.\ (\ref{RWmetric1}) is usually referred to as the 
\emph{Friedmann equation}. Together with Eq.\ (\ref{RWmetric2}), it implies
\begin{equation}\label{firstlaw}
\dot{\rho} \: + \: 3 \left( \rho \: + \: P \right) \frac{\dot{a}}{a} \; = \; 0
\end{equation}
(alternatively, Eq.\ (\ref{firstlaw}) is the component of the equation of motion of the fluid, $\nabla^a T_{ab} = 0$,
parallel to the congruence).
\end{itemize}
In order to determine the dynamical behaviour of the scale factor, $a$, as a function of pressure, $P$, and energy density, $\rho$,
of the cosmic fluid, it is necessary to posit an \emph{equation of state}, i.e., a functional relationship between
$P$ and $\rho$, that allows elimination of $P$ from Eq.\ (\ref{firstlaw}) and subsequent expression of $\rho$ as a
function of $a$. Usually such an equation of state is simply assumed to be linear, i.e.,
\begin{equation}
P \; = \; w \rho 
\end{equation}
for a constant proportionality factor, $w$, which in the following is taken to be non-negative\footnote{Although all
of the following discussion goes through unchanged by allowing negative values for $w$ such that $w > - 1/3$, 
pressure is, at the present, classical level of discussion, taken to be a non-negative quantity on physical grounds 
(similar remarks pertain to the classical energy density of matter, $\rho$).
In particular, the matter distribution at the present epoch effectively behaves as ``dust'' ($w=0$), whereas in
the early Universe, it effectively behaved as ``radiation'' ($w=1/3$). No negative-pressure classical matter is
currently known to exist.
Although the cosmological constant, $\Lambda$, is sometimes interpreted as a negative-pressure, cosmic perfect fluid (with
$w=-1$), for various reasons this view will not be adopted here. In accordance with Eq.\ (\ref{Einstein}),
$\Lambda$ is interpreted as being associated with ``geometry'', rather than with ``sources'', in the current work.
Nevertheless, it will occasionally prove useful to temporarily switch views in order to facilitate comparison with existing literature.
Finally, it is clear that the present discussion can be generalized to the case of non-interacting, multi-component perfect fluids
(consisting e.g.\ of both radiation and dust simultaneously; see also note \ref{BBinevitable}).}. Integration of Eq.\ (\ref{firstlaw}) then gives 
\begin{equation}\label{rhostate}
\rho \; = \; C a^{-3(1 + w)}
\end{equation}
with $C \geq 0$ an arbitrary constant and the Friedmann equation becomes
\begin{equation}\label{Friedmanneq}
\dot{a}^2 \; = \; \frac{8 \pi}{3} C a^{-(1 + 3w)} \: + \: \frac{\Lambda a^2}{3} \: - \: k
\end{equation}
General solutions to Eq.\ (\ref{Friedmanneq}) for various cases of physical interest can be written down explicitly
(some in terms of elementary functions), but it suffices here to restrict attention to the two most important
generic features of solutions in relation to the present discussion\footnote{For a fairly comprehensive 
treatment of the various possible FLRW models, see for instance Ref. \cite{Rindler}.\label{FLRWfeatures}
For specific treatments from a dynamical systems perspective, see Ref. \cite{StRe,*WaEl,*UzLe}.
The value of the cosmological constant turns out to play a critical role in the dynamical behaviour of solutions. In particular, 
\emph{all} models with $\Lambda < 0$ are ``oscillatory'' (i.e., reach a stage of maximum expansion and then re-contract), while \emph{all} models with $\Lambda > \Lambda_{\mathrm{E}}$,
i.e., $\Lambda$ greater than the critical value, $\Lambda_{\mathrm{E}} := 4 \pi (1 + 3w) \rho$, corresponding to a general Einstein static universe,
are forever expanding (so-called \emph{inflexional universes}).
For non-negative values of $\Lambda$ not greater than $\Lambda_{\mathrm{E}}$, there is a difference between models according to whether
their spatial geometries have positive curvature or not: the ``standard'' models with $k=+1$ behave in the same way as their
counterparts with negative $\Lambda$, i.e., they are oscillatory, while models with $k=-1$ and $k=0$ are forever expanding.
The exceptions for the positive curvature models arise for $0 < \Lambda < \Lambda_{\mathrm{E}}$ in the form of ``rebounding universes''
(which start from a state of infinite expansion to reach a minimum radius and then re-expand) and for $\Lambda = \Lambda_{\mathrm{E}}$
in the form of two non-static solutions, which respectively start from and asymptotically tend to the static solution (the
former is sometimes regarded as a perturbed Einstein static universe and in the specific case of dust, $w=0$, it is then referred to as the \emph{Eddington-Lema\^itre universe}).
As will be seen in subsection \ref{instable}, observations at present indicate that if in fact the Universe's geometry is elliptic, $\Lambda$ is significantly larger than $\Lambda_{\mathrm{E}}$.
Finally, the singularity at $a=0$ discussed in the main text (conventionally taken to occur at the ``initial time'' $\tau = 0$) is characterized by the following features:
(i) zero distance between all points of space, (ii) infinite density of matter and (iii) infinite curvature for 
$k \neq 0$ (i.e., for small $\tau$, the scalar curvature, $R$, tends to $6k/a^2$).}: 
\begin{itemize}
\item[1.] All \emph{physically relevant} FLRW models - i.e., models that are (at least at some stage) effectively ``non-empty'' 
($C > 0$) and have sufficiently large, positive cosmological constant, $\Lambda$ - are \emph{non-static} and start with a \emph{``Big-Bang''}, i.e., a singularity, $a=0$, at $\tau=0$
\item[2.] For small $\tau$, the dynamical trajectories of \emph{all} FLRW models with an initial singularity tend to the corresponding trajectories of $k=0$, $\Lambda=0$ models (as can in fact
be read off directly from the Friedmann equation in the form (\ref{Friedmanneq})). In particular, regardless of spatial 
geometry and the value of the cosmological constant, at early times all past-singular solutions for dust tend to $\tau^{2/3}$ 
(Einstein-de Sitter solution), whereas all past-singular solutions for radiation tend to $\tau^{1/2}$
\end{itemize}
As will be seen in the next section, both these features play a critical role in what is widely perceived to be a 
serious shortcoming of FLRW models as applied to the actual Universe.

\section{Problems of Flatness - Inflated and Otherwise}\label{FP}

\noindent Fundamental to most discussions of the putative flatness problem(s) is the Friedmann equation (\ref{RWmetric1}) 
rewritten in the following form
\begin{equation}\label{Friedmanneq2}
\Omega_{\mathrm{m}} \: + \: \Omega_{\Lambda} \; = \; 1 \: + \: \frac{k}{\dot{a}^2} \; = \;1 \: + \: \frac{k}{H^2 a^2} 
\end{equation}
where the dimensionless parameters $\Omega_{\mathrm{m}} := 8   \pi   \rho / 3H^2$ and $\Omega_{\Lambda} := \Lambda / 3H^2$ 
respectively measure the density of matter, $\rho$, and the (suitably normalized) cosmological constant, 
$\Lambda/8 \pi$, in terms of the ``critical density'', $\rho_{\mathrm{c}} := 3H^2 / 8 \pi$ (see below),  and where $H := \dot{a} / a$ denotes Hubble's ``constant''. 
Historically, the line of argument that was taken to establish a ``flatness problem'' first emerged in the late 1970s, when (in
absence of empirical evidence to the contrary) the cosmological constant was typically taken to vanish and for simplicity
this context will be dealt with first in subsection \ref{finetune} below (as will also become clear however, inclusion of nonzero $\Lambda$ 
does not change anything as far as this particular line of argument is concerned). On setting $\Lambda=0$, it is immediately clear from 
Eq.\ (\ref{Friedmanneq2}) that $\rho_{\mathrm{c}} $ is indeed a critical density required for an ``open universe''. 
That is, a FLRW model is necessarily ``closed'' (spatial geometry of finite extension, corresponding to $k=+1$) 
if $\rho > \rho_{\mathrm{c}}$, whereas it is usually called ``open'' (spatial geometry of infinite extension, corresponding to 
$k=-1$ or $k=0$)\footnote{As observed earlier (cf. note \ref{topology}), both hyperbolic and Euclidean geometries
admit topologically closed models, although these appear to be somewhat contrived.
It should also be noted that there is a potential ambiguity in the open/closed universe terminology, as it might seem that 
reference could be made to either spatial geometry (as is intended here) or \emph{temporal duration}.
In fact, in the present context with $\Lambda = 0$, the distinction is irrelevant as long as the standard practice
of referring to negative and zero curvature models as ``open'' is followed.
This, however, ceases to be the case for nonzero $\Lambda$. 
In particular (cf. note \ref{FLRWfeatures}), all models with open spatial geometry ($k=-1, 0$) are temporally closed (i.e., eventually recontract) 
if $\Lambda <0$, while all models with closed spatial geometry ($k=+1$) are temporally open (i.e., expand indefinitely) if $\Lambda > \Lambda_{\mathrm{E}}$.} if $\rho \leq \rho_{\mathrm{c}}$.
Moreover, unlike the possible models with nonzero curvature, the $\Omega$-parameter for the critical case, corresponding
to a flat, Euclidean spatial geometry, is identically constant, i.e., $\Omega_{\mathrm{m}}=1$ at all times.
Now, the actual observed value of $\Omega_{\mathrm{m}}$ at present is known to be close to unity \cite{Adecs} - i.e., close to
the constant value corresponding to a Euclidean geometry - and the flatness problem of FLRW cosmologies, broadly construed,
is perceived to be the apparent ``improbability'' of this basic empirical truth.\\
In fact, a critical inspection of the relevant literature indicates that there are at least three different formulations
of the flatness problem and although these formulations are related to each other (and not infrequently also conflated), they 
are ultimately distinct.
A basic classification of (alleged) flatness problems can be made depending on whether \emph{dynamical} arguments are
involved or not. Those that do are found to exist in two standard versions (which, somewhat ironically, are logically complementary
to each other), namely (i) a perceived problem  of exceeding unlikeliness that $|\Omega_{\mathrm{m}} - 1|$, if in fact nonzero, was very nearly ``almost'' zero -
i.e., to within a very large number of decimal places - at very early times, \emph{given that} $\Omega_{\mathrm{m}} \simeq 1$ 
at present (fine-tuning argument) and (ii) a perceived problem of exceeding unlikeliness that, in the non-critical case, $\Omega_{\mathrm{m}} \simeq 1$ at 
present, \emph{given that} $|\Omega_{\mathrm{m}} - 1|$ was ``extremely close'' to zero at very early times (instability argument).
Non-dynamical versions of the flatness problem on the other hand are essentially variations on the theme of why 
fundamental physical parameters, such as Newton's constant, $G_{\mbox{\scriptsize N}}$, or the fine-structure constant, $\alpha$, 
take the particular values they do.

\subsection{The Fine-Tuning Argument}\label{finetune}

\noindent According to the prevailing wisdom, in order that the value of $\Omega_{\mathrm{m}}$ be anywhere near $1$ today, it had to 
be extremely delicately ``adjusted'' in the very early Universe. There is then an arguable necessity to identify the 
relevant physical laws responsible for such an alleged ``fine-tuning'' and it is often argued in this 
regard that ``inflationary physics'' has the potential to provide a resolution (see section \ref{IC}).
The essential steps of the argument are to derive from the Friedmann equation (\ref{Friedmanneq2}) the following identity
\begin{equation}\label{Friedmanneq4}
\left( \Omega_{\mathrm{m}} (\tau') \: - \: 1 \right) \; = \; \left( \Omega_{\mathrm{m}} (\tau) \: - \: 1 \right) \frac{\dot{a}^2_{\tau}}{\dot{a}^2_{\tau'}} 
\end{equation}
(where $\tau$ and $\tau'$ explicitly display the proper time dependence of quantities but are further arbitrary) and to note that 
Eq.\ (\ref{RWmetric2}) (in the present context with $\Lambda = 0$, $C > 0$ and $w \geq 0$) implies
\begin{equation}
\ddot{a} \; < \; 0 
\end{equation}
at all times. Since $\dot{a}$ blows up at the origin, this means that for fixed $\tau$ and sufficiently small $\tau'$, one 
has $\dot{a}^2_{\tau}/\dot{a}^2_{\tau'} \ll 1$.
Thus, if $\tau$ denotes the present epoch, for which $\Omega_{\mathrm{m}}$ differs from
$1$ by, say, a number of order unity, it directly follows from Eq.\ (\ref{Friedmanneq4}) that the deviation of $\Omega_{\mathrm{m}}$
from $1$ in the very early Universe had to be ``extremely small''
\begin{equation}
|\Omega_{\mathrm{m}} (\tau') \: - \: 1 | \; \ll \; 1
\end{equation}
so that (according to this argument) $\Omega_{\mathrm{m}} (\tau')$ had to be seriously ``fine-tuned'' in some way.

\subsubsection*{Some Numerology} 

\noindent It is instructive to consider a few standard examples with explicit, numerical estimates of the amount of 
putative fine-tuning of $\Omega_{\mathrm{m}}$. 
First, since $\Omega_{\mathrm{m}}$ is presently close to $1$, it is (tacitly) assumed that the $k=0$ solutions provide a good approximation.
Taking this assumption to be valid (if only for the sake of argument) and using the
explicit solutions for dust (i.e., $a \sim \tau^{2/3}$) and radiation (i.e., $a \sim \tau^{1/2}$) in this case,
Eq.\ (\ref{Friedmanneq4}) rewrites to
\begin{equation}\label{Friedmanneq5}
\left( \Omega'_m \: - \: 1 \right) \; = \; \left( \Omega_{\mathrm{m}} \: - \: 1 \right) \left( \frac{\tau'}{\tau} \right)^{2/3} \qquad \mbox{and} \qquad \left( \Omega'_m \: - \: 1 \right) \; = \; \left( \Omega_{\mathrm{m}} \: - \: 1 \right) \frac{\tau'}{\tau}
\end{equation}
respectively. Taking $\tau \simeq 13.8 \, \mbox{Gy} \simeq 10^{17} \, \mbox{s}$, $\tau_{\mbox{\scriptsize eq}} \simeq 40 \, \mbox{ky} \simeq 10^{12} \, \mbox{s}$ and $\Omega_{\mathrm{m}} \simeq 0.3$ 
as representative values for respectively the current age of the Universe, the age of the Universe when matter started to
become dominant over radiation and the (normalized) density of matter corresponding to the current 
epoch, this then leads to a ``fine-tuning'' of $\Omega_{\mathrm{m}}$ to $1$ to an accuracy of 
\begin{equation}\label{finetuneexp1}
\mbox{one part in} \; 10^{16}
\end{equation}
if $\tau'$ is taken to be of the order of a second. Similarly, $\Omega_{\mathrm{m}}$ is ``fine-tuned'' to $1$ to an accuracy of
\begin{equation}\label{finetuneexp2}
\mbox{one part in} \; 10^{59}
\end{equation}
if $\tau'$ is taken to be of the order of the Planck time $\simeq 10^{-43} \, \mbox{s}$.

\vspace{0.5cm}

\noindent As just presented, the fine-tuning argument is the most commonly encountered line of reasoning - often also the only one considered - purported to establish
a flatness problem. Going back to an early argument (made somewhat less explicitly) by Dicke and Peebles \cite{DiPe},
it was subsequently emphasized by Guth and others as a key motivation for introducing inflationary models \cite{Guth1,*Linde3,*Albstein}
and has been viewed as signalling a major drawback of the standard FLRW models by probably most cosmologists ever since\footnote{Explicit
arguments leading to the sort of numerical estimates given in the main text can be found in many references (see Ref. \cite{Guth3,*Guth4,*Linde,*Ryden,*Baumann,*HaHo}
for a fragmentary list of relevant arguments).
It should be noted that slightly different estimates of the alleged fine-tuning can be found in the actual literature, depending
for instance on what value for $\tau_{\mbox{\scriptsize eq}}$ is adopted or on the exact nature of the calculation (in
some older references estimates can be found that are based on calculations also involving quantum statistical arguments and
entropy bounds - leading, somewhat miraculously, to similar numerical estimates; more modern references typically
follow the line of argument presented in the main text, being fundamentally based only on classical FLRW dynamics).
However, it appears that the figure of $59$ decimal places for a Planck order cut-off time may be taken 
as some sort of weighted average.
A fundamental problem with all these approaches that is hardly ever pointed out however, is that the step of taking
flat solutions as valid approximations to all viable solutions for the entire history of the Universe, of course completely
begs the question. In fact, Eqs.\ (\ref{Friedmanneq5}) are somewhat misleading, since for flat solutions one obviously has $\Omega_{\mathrm{m}} = 1$ for all times!
But since there is no guarantee that a curved solution is in any reasonable sense near its flat counterpart if $\Omega_{\mathrm{m}}$ is 
actually close to $1$, \emph{except} at very small times (as the cycloidal model discussed in subsection \ref{instable} for 
instance perfectly well illustrates), the actual figures quoted, (\ref{finetuneexp1}), (\ref{finetuneexp2}) are rather deceptive as things stand.\label{finetunecaveat}}. 
It is however a seriously problematic argument - as was in fact already pointed out in different ways by various authors 
on previous occasions \cite{CoEl}, \cite{KiEl}, \cite{AdOv}, \cite{Lake}, \cite{Helbig,*Helbig2}, \cite{Carroll}.
The essential point is that the near-magical ``fine-tuning'' exhibited by Eqs.\ (\ref{finetuneexp1}), (\ref{finetuneexp2}),
is nothing but a \emph{consequence} of the Friedmann dynamics (\ref{Friedmanneq2}) (or, equivalently, Eq.\ (\ref{RWmetric1})).
Indeed, as stressed in the previous section (cf. the discussion below Eq.\ (\ref{Friedmanneq})), \emph{all} cosmologically relevant
FLRW models \emph{necessarily} become indistinguishable from flat FLRW models in the limit of small $\tau$, in the sense that $\Omega_{\mathrm{m}} = 1$ exactly at $\tau = 0$.
This for instance follows directly from Eq.\ (\ref{Friedmanneq2}), since $\dot{a}$ diverges for $\tau \rightarrow 0$; alternatively,
it follows from expressing $\Omega_{\mathrm{m}}$ in terms of the scale factor, $a$, using Eqs.\ (\ref{rhostate}), (\ref{Friedmanneq}), i.e.,
\begin{equation}\label{Omegamatter}
\Omega_{\mathrm{m}} \; = \; \frac{8 \pi C}{8 \pi C - 3 k a^{1 + 3w}}
\end{equation}
In other words, the ``fine-tuning'' is a feature that is \emph{inherent} to all these models. In particular, it also occurs
in those instances where $\Omega_{\mathrm{m}}$ is \emph{not} close to $1$ after $\simeq 10^{17} \, \mbox{s}$ (for instance,
it is easily checked that solutions for which $\Omega_{\mathrm{m}}$ has become significantly smaller than $1$ after this
time are ``fine-tuned'' to the same fantastic precision - i.e., to $59$ decimal places in the above example - at the Planck scale in this sense just as well; 
clearly $|\Omega_{\mathrm{m}} - 1|_{\mbox{\scriptsize present}}$
is of order $1$ for all values of $\Omega_{\mathrm{m}}$ of order $10^{-N}$, $N \geq 0$).
There is thus no problem of fantastic improbability, since initial conditions \emph{could not have been otherwise} within
the present context.\\ 
It is useful to slightly elaborate upon this point, since it is sometimes contended that $\Omega_{\mathrm{m}}$ could have had \emph{any}
initial value (see some of the earlier quoted references for explicit statements to this effect).
As the foregoing shows, within the strict context of isotropic cosmologies, this contention is simply false.
Although actually underlying the claim is a different context (which is usually not very clearly spelled out), the conclusion that there is no fine-tuning problem is unaffected.
First, as various authors have pointed out, the assumption that the set of possible initial values for $\Omega_{\mathrm{m}}$ is significantly larger and governed by a sufficiently
uniform probability distribution can well be criticized on general methodological grounds \cite{CoEl}, \cite{Helbig,*Helbig2}, \cite{EvCo}.
In fact, such a view is also supported by a sober analysis of a more explicit picture of initial conditions.\\
That is, even though a consistent, physical theory of ``quantum gravity'' is yet to be formulated, it is often argued that the classical picture of spacetime as described 
by general relativity should reasonably be expected to break down near the Planck scale.
If it is assumed that the FLRW context provides an acceptable description of spacetime geometry all the way down to the Planck time $\sim 10^{-43} \, \mbox{s}$ - which is what 
most of those who argue for an actual independent flatness problem seem perfectly content with (given their numerological arguments at least) - and if the set of possible ``initial'' 
values for $\Omega_{\mathrm{m}}$ at this time is allowed to be very general, say $\real{+}$, it is certainly true that $|\Omega_{\mathrm{m}} - 1| \simeq 10^{-59}$ (or some similar 
small number) as an initial condition appears extremely unnatural at first glance.
What is often not realized however is that there is a sharp ``trading off'' between genericity in an initial value of $\Omega_{\mathrm{m}}$ at the Planck scale and genericity in 
the parameter $C$ residing in $\Omega_{\mathrm{m}}$ (and similarly for other FLRW models with different matter contents).
The reason is simply that FLRW solutions, in order to prevent their duplication, display a one-to-one correspondence between possible ``initial'' values of $\Omega_{\mathrm{m}}$
at the Planck scale $\sim 10^{-43} \, \mbox{s}$ (i.e., $0 < \Omega_{\mathrm{m}} < 1$ for $k =-1$ and $1 < \Omega_{\mathrm{m}} < \infty$ for $k =+1$) and possible values of $C$.
In particular, the more generic an initial value for $\Omega_{\mathrm{m}}$, the more fine-tuning in $C$ would be required to achieve this.
As can readily be verified from Eq.\ (\ref{Omegamatter}), for \emph{generic} values of $C$ (as picked at random, say, by
an unskilled, blindfolded Creator), $\Omega_{\mathrm{m}}$ will still be fine-tuned to the fantastic number of decimal places (\ref{finetuneexp2}) at the Planck scale.\\
The fact that spatial geometry could be highly anisotropic at pre-Planckian timescales - as seems to be envisaged in many currently popular approaches - is irrelevant here.
Although it is prima facie unclear how to extend the notion of a (single) ``critical density'' - and hence the definition of $\Omega_{\mathrm{m}}$ - in a meaningful way to 
general anisotropic models (for homogeneous models it is still possible to define an average Hubble constant \cite{Misner}), if a more general set of initial values for 
$\Omega_{\mathrm{m}}$ is envisaged to (somehow) carry physical meaning, the enormous ``fine-tuning'' of $\Omega_{\mathrm{m}}$ is a mere artefact of the particular construction at hand
and would in itself thus establish nothing.
This is because any (cosmologically relevant) model for which $\Omega_{\mathrm{m}}$ is \emph{not} so ``fine-tuned'' cannot be a FLRW model and hence, the tacit underlying issue here
is that of natural probability assignment to FLRW models in some ``sufficiently large space of theoretically consistent cosmological models'' (and obviously, the exact same argument 
can be made for models that are ``almost isotropic''; these would also have to be sufficiently ``fine-tuned'').\\
As will be discussed further in section \ref{IC}, although there are indeed good grounds to argue that FLRW models are in some
sense ``improbable'', these grounds, being fundamentally based on the second law of thermodynamics, are rather different
from the type of probabilistic reasoning that is usually encountered in this context. In fact, as will become clear from this 
particular perspective on things, it is highly unlikely that the improbability of FLRW initial conditions in the sense
intended, can be succesfully resolved merely in terms of dynamical arguments (as invariably called upon in response to
the more usual kind of improbability claims).
But whichever particular perspective is adopted, all probabilistic arguments relevant here merely address the 
overall context of isotropy. Once within that context, a ``fine-tuning'' of $\Omega_{\mathrm{m}}$ and the existence
of particle horizons (see section \ref{IC}) are inevitable features of all cosmologically relevant models.
This thus means that there is no independent problem of flatness of the fine-tuning variety.\\
As is easily verified, this conclusion is completely general, despite the fact that the cosmological constant, $\Lambda$,
was ignored in the foregoing. Using again Eq.\ (\ref{Friedmanneq}), the parameter $\Omega_{\Lambda}$ of Eq.\ (\ref{Friedmanneq2})
can be expressed directly in terms of the scale factor
\begin{equation}\label{Omegalambda}
\Omega_{\Lambda} \; := \; \frac{\Lambda}{3H^2} \; = \; \frac{\Lambda a^{3(1 + w)}}{8 \pi C + \Lambda a^{3(1 + w)} - 3k a^{1 + 3w}} 
\end{equation}
from which it follows that $\Omega_{\Lambda} \rightarrow 0$ for $a \rightarrow 0$ (for $\Omega_{\mathrm{m}}$, an additional
term $\Lambda a^{3(1 + w)}$ now also has to be included in the denominator of expression (\ref{Omegamatter}), but this term
obviously does not affect the limiting behaviour of $\Omega_{\mathrm{m}}$ at $a=0$).
Thus, if near-flatness is understood in the (physically more appropriate) sense of ``closeness to $1$'' of the total $\Omega$-parameter,
i.e., $\Omega := \Omega_{\mathrm{m}} + \Omega_{\Lambda}$, rather than in terms of the approximate equality of the density of matter
to the critical density, the same calculation as before now gives that $\Omega$ differed from $1$ by a number
of order $10^{-62}$ at the Planck scale, if its present deviation from $1$ is $\lesssim 10^{-3}$ \cite{Adecs}.
Inclusion of an actual, physically relevant cosmological constant term only increases slightly the \emph{level} of
putative fine-tuning; it does not in any way affect the previous conclusion that there is no independent fine-tuning
problem to begin with.

\subsection{The Instability Argument}\label{instable}

\noindent A more sophisticated argument purporting to establish the problematic nature of the near-flatness of the Universe's 
large-scale spatial geometry essentially consists of the claim that \emph{given} that (to a fantastic number of decimal places)
the initial density of matter, $\rho$, equaled the critical density, $\rho_{\mathrm{c}}$, or, alternatively, that the initial value of (total) $\Omega$ was equal to $1$, it is 
not to be ``expected'' that (within an order of magnitude, say) $\rho$ would still be ``close'' to $\rho_{\mathrm{c}}$, or, alternatively,
that $\Omega \simeq 1$, after some fourteen or so billion years of cosmic evolution (see e.g.\ Ref. \cite{Weinberg,*Smeenk,*SmEl} for an argument along these lines).
Given that $\rho$ necessarily evolves away from $\rho_{\mathrm{c}}$ with time (for nonzero $k$), i.e., $\Omega = 1$
is an unstable fixed point of the FLRW dynamics, this seems a prima facie plausible argument (although, as already
mentioned, different types of arguments pertaining to flatness are typically conflated in the literature - something 
which is in particular true for the instability and fine-tuning varieties).
Upon closer inspection however, this argument is also found to evaporate into thin air.
First, if the present closeness of $\Omega$ to $1$ is accepted as basic empirical input and the approximations
used in section \ref{finetune} are again taken to be valid, it is clear that there can be no probability problem 
in a restricted sense, since the (deterministic) FLRW dynamics works both ways.
In order for this type of argument to be potentially successful, it is necessary to consider a significantly more general framework,
in which models where $\Omega$ remains ``close to $1$'' for ``sufficiently long'' either do not exist or can somehow 
be securely classified as being highly non-generic. This however does not appear to be possible. (Note that even if the instability argument
were sound, a dynamical mechanism, such as inflation, designed to drive spatial geometry to near-flatness at extremely early times, would by its very nature be useless 
to address this argument.)\\
For instance, for a cycloidal universe (i.e., a dust-filled model with positive curvature and zero cosmological constant; 
$w=\Lambda=0$, $k=+1$), it is easily checked that $\Omega_{\mathrm{m}} < 10$ for a full $60 \%$ of its total lifespan, even though $\Omega_{\mathrm{m}}$ is in fact unbounded.
Fictitious theorists in such a hypothetical universe would thus presumably conclude that there was nothing improbable
to explain if observations were in fact to indicate that $\Omega_{\mathrm{m}} < 10$, since probabilities for observing different 
ranges of $\Omega_{\mathrm{m}}$-values are evidently distributed in a highly non-uniform fashion.\\
In view of the actual constraints placed on the present value of $\Omega_{\mathrm{m}}$ by observations (which exclude the
possibility that $1 < \Omega_{\mathrm{m}} < 10$) and, more generally, in view of the oscillatory nature of the cycloidal universe,
the example just presented is of course rather unrealistic on physical grounds\footnote{It is recalled that all oscillatory
models (i.e., all models with $\Lambda < 0$, regardless of the value of $k$, and all models with $0 \leq \Lambda < \Lambda_{\mathrm{E}}$
and $k=+1$) are dynamically constrained to the region to the left of or below the locus of $\ddot{a}=0$
in the $a \Lambda$-plane and, as a consequence, have $\ddot{a}<0$ at all times, whereas observations indicate that 
the universal expansion is accelerating, i.e., that $\ddot{a}>0$ at present. 
In fact, in terms of the current value of the ``deceleration parameter'', $q := - a \ddot{a} / \dot{a}^2$,
one has (cf. Eq.\ (\ref{RWmetric2})) $q = \Omega_{\mathrm{m}}/2 - \Omega_{\Lambda} \simeq - 0.55$.}.
Yet, it serves well to illustrate the point that something which may seem a priori obvious (i.e., the idea that the 
cycloidal universe would be ``far more probable'' to reside in the infinite region $10 < \Omega_{\mathrm{m}} < \infty$, 
than in the finite region $1 < \Omega_{\mathrm{m}} < 10$, both regions being traversed twice), need not in fact be true.
Furthermore, a similar line of argument can be used to establish that there is no manifest probability problem in the
non-oscillatory case either (see Ref. \cite{Rindler} for a discussion of both the cycloidal and non-cycloidal cases),
something which is intuitively obvious from the expression for the total $\Omega$-parameter
\begin{equation}\label{Omegatotal}
\Omega \; := \; \Omega_{\mathrm{m}} \: + \: \Omega_{\Lambda} \; = \; \frac{8 \pi C + \Lambda a^{3(1 + w)}}{8 \pi C + \Lambda a^{3(1 + w)} - 3k a^{1 + 3w}} 
\end{equation}
Indeed, it is clear from this expression that for cosmologically relevant models, $\Omega=1$ is both a repulsor (i.e.,
at $a=0$) and an attractor (i.e., for $a \rightarrow \infty)$ of the dynamics and it is therefore not obviously ``strange'' or 
improbable if the value of $\Omega$ should happen to be observed near $1$ at some intermediate time, e.g.\ at $13.8 \, \mbox{Gy}$.
In fact, a rigorous analysis of possible dynamical trajectories in the $\Omega_{\mathrm{m}} \Omega_{\Lambda}$-plane, taking into
account actual observational constraints on these parameters from precision measurements of the CMB and of type IA 
supernovae, fully confirms this intuitive argument. To this end, it is useful to introduce a new, dimensionless time parameter according to $\eta := \ln (a/a_0)$,
$a_0$ denoting some fixed reference value for the scale factor, here taken to correspond to the present epoch.
This thus means that $' := d/d\eta = H^{-1} d/d\tau$ and on taking the case of dust ($w=0$) to be representative, Eqs.\ (\ref{RWmetric1}), (\ref{RWmetric2})
imply that $(\Omega_{\mathrm{m}},\Omega_{\Lambda})$ satisfy the following autonomous system of first-order differential equations
\begin{eqnarray}
\Omega_{\mathrm{m}}'	       & = & (\Omega_{\mathrm{m}} - 2 \Omega_{\Lambda} - 1)\Omega_{\mathrm{m}} 		   \label{FLRWdyn1}  \\
\Omega_{\Lambda}'      & = & (\Omega_{\mathrm{m}} - 2 \Omega_{\Lambda} + 2)\Omega_{\Lambda}           \label{FLRWdyn2}
\end{eqnarray}
Conversely, Eqs.\ (\ref{FLRWdyn1}), (\ref{FLRWdyn2}) imply Eqs.\ (\ref{RWmetric1}), (\ref{RWmetric2}) for $P=0$ and hence the system (\ref{FLRWdyn1}), (\ref{FLRWdyn2})
constitutes an equivalent specification of the FLRW dynamics in the case of dust.
Nontrivial critical points of (\ref{FLRWdyn1}), (\ref{FLRWdyn2}) are a repulsor (or negative attractor) at $(\Omega_{\mathrm{m}}, \Omega_{\Lambda})=(1,0)$,
corresponding to the Big Bang and a (positive) attractor at $(\Omega_{\mathrm{m}}, \Omega_{\Lambda})=(0,1)$, corresponding to an asymptotic de Sitter phase.
It is straightforward to check that the function $\alpha$, defined by
\begin{equation}\label{FLRWconstant}
\alpha \; = \; \mp \frac{27 \Omega^2_{\mathrm{m}} \Omega_{\Lambda}}{4 \Omega_k^3} \qquad \quad \mbox{for} \; \; k = \pm 1, \quad \Omega_{\mathrm{m}} \geq 0 , \quad \Omega_{\Lambda} \geq 0 , \quad \Omega_k := -k/\dot{a}^2
\end{equation}
is a first integral of the system (\ref{FLRWdyn1}), (\ref{FLRWdyn2}) (note that for any FLRW solution, $\Omega_k = 1 - \Omega_{\mathrm{m}} - \Omega_{\Lambda}$). That is, the derivative of $\alpha$ along any orbit of (\ref{FLRWdyn1}), (\ref{FLRWdyn2}) vanishes
\begin{equation}
\alpha ' \; = \; \frac{\partial \alpha}{\partial \Omega_{\mathrm{m}}} \Omega_{\mathrm{m}}' \: + \: \frac{\partial \alpha}{\partial \Omega_{\Lambda}} \Omega_{\Lambda}' \; = \; 0
\end{equation}
and $\alpha$ is thus a constant of the motion\footnote{For $k=+1$, $\Lambda >0$ models, the physical interpretation of $\alpha$ is that of
the cosmological constant, $\Lambda$, normalized with respect to the critical value $\Lambda_{\mathrm{E}}$ corresponding to an
Einstein static universe, i.e., $\alpha = \Lambda / \Lambda_{\mathrm{E}}$ (alternatively, $\alpha \sim \Lambda M^2$, with $M$ denoting the total mass of the universe).
As follows from the discussion in the main text, this justifies a remark made earlier, that for physically relevant elliptic FLRW models $\Lambda \gg \Lambda_{\mathrm{E}}$.}.
\begin{figure}
\begin{center}
\includegraphics[trim={20mm 10mm 10mm 60mm},width=100mm]{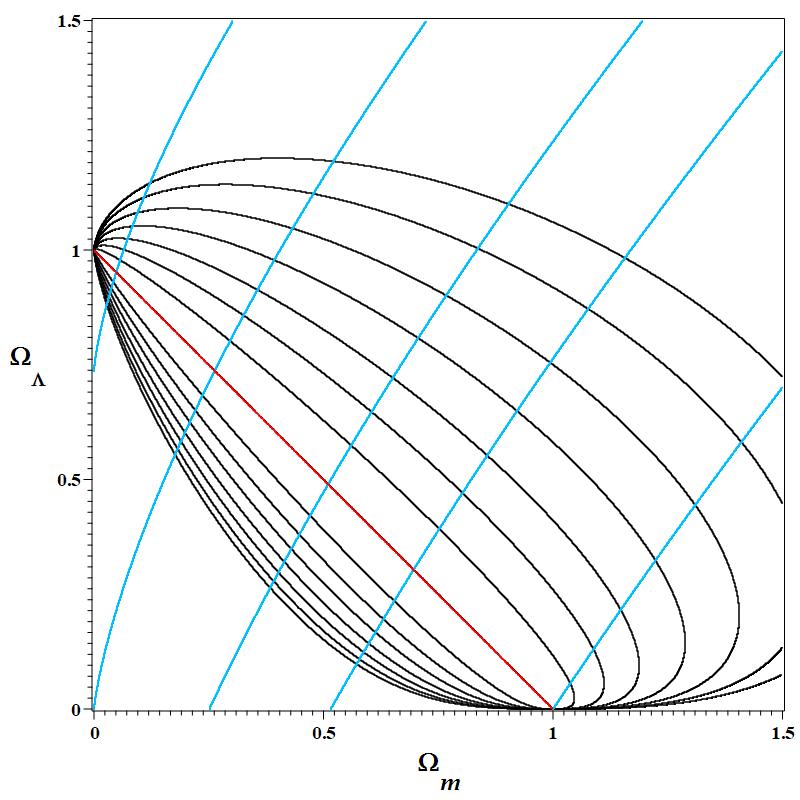}
\caption{Partial Phase Portrait of $w=0$ FLRW Dynamical Trajectories.
The straight line segment represents a flat FLRW model and separates the hyperbolic ($k=-1$) and elliptic ($k=+1$) models within the phase plane (which respectively
lie to the left and to the right of the $k=0$ model as it is traversed into the future).
Depicted in black are fourteen level sets of $\alpha$ that approach the delimiter from below, respectively above, for the subsequent values $\alpha = 6, 8, 12, 20, 40, 100, 500$.
The delimiter itself corresponds to $\alpha = \infty$.
The dynamical trajectories are traversed by level sets of $tH$, here shown (in blue), from right to left, for the values $tH = 2/3, 3/4, 0.826, 1, 3/2$.
The particular values for $\alpha$ and $tH$ are partially informed by the results from WMAP and HST (cf. Ref. \protect\cite{Lake}).
Level sets of $\alpha$ corresponding to lower bound values in accordance with the more recent Planck data have not been included, since these are near-indistinguishable from the $\alpha = \infty$ delimiter.}\label{FLRWtraj}
\end{center}
\end{figure}
Fig. \ref{FLRWtraj} depicts a family of level sets $\alpha = \mbox{constant}$ in a partial phase plane, intersected by a number of level sets of $tH$, where $t$ denotes the
age of the Universe\footnote{Eq.\ (\ref{ageuniverse}) follows from the obvious fact that for any reference epoch, $t_0 H_0$ is given by 
\begin{equation}
t_0 H_0 \; = \; \int_0^{a_0} \frac{da}{aH/H_{0}} \; = \; \int_0^1 \frac{dx}{x E(x)} \qquad \qquad x \; := \; a/a_0 \; = \; e^{\eta} 
\end{equation}
where $E \; := \; H/H_0 = \sqrt{\Omega_{\mathrm{m},0} x^{-3} + \Omega_{\Lambda,0} + \Omega_{k,0} x^{-2}}$.
The reference subscript is implicit in (\ref{ageuniverse}).
Alternatively, 
\begin{equation}
t \; = \; \frac{1}{H} \, \int^{\infty}_0 \frac{dz}{(1 + z)\sqrt{(1 + z)^2(1 + z\Omega_{\mathrm{m}}) - z(z + 2) \Omega_{\Lambda}}}
\end{equation}
where $z$ denotes the usual redshift factor, i.e., $(1+z)^{-1} = e^{\eta} = a/a_0$.}
\begin{equation}\label{ageuniverse}
t \; = \; \frac{1}{H} \, \int^1_0 \frac{dx}{\sqrt{\Omega_{\mathrm{m}} x^{-1} + \Omega_{\Lambda}x^2 + 1 - \Omega_{\mathrm{m}} - \Omega_{\Lambda}}}
\end{equation}
The straight line segment $\Omega_{\Lambda} = 1 - \Omega_{\mathrm{m}}$, representing a flat FLRW model, is asymptotically approached from below, resp. above, by the hyperbolic (i.e., $k=-1$),
resp. elliptic (i.e., $k=+1$) FLRW models in the large-$\alpha$ limit.
The significance of these remarks is that CMB precision measurements strongly suggest that the $w=0$ FLRW model that best describes our Universe in its current phase is one
for which $\alpha$ is very large.
To the extent that such a model can be taken as a good approximation for the Universe during its \emph{entire} evolution, it is then clear that there is no probability/instability
problem, since such a model has $\Omega$ (as defined by Eq.\ (\ref{Omegatotal})) close to unity throughout its entire history.
As will be seen shortly however, this conclusion is in fact of much more general validity and in particular also pertains to FLRW models that went through a transition from a 
radiation-dominated epoch at early times, to a matter-dominated epoch (in the form of dust) at late times - as is estimated to have occurred in our actual Universe, about $380,\!000$ years after the Big Bang.\\
The above argument, linking the large value of the FLRW constant of the motion (\ref{FLRWconstant}) implied by cosmological observations, to the flatness issue, first appears to have been 
made by Lake over a decade ago \cite{Lake}.
At that time, the results from the Wilkinson Microwave Anisotropy Probe (WMAP) and the Hubble Space Telescope (HST) on respectively CMB anisotropies and type IA supernovae,
constrained curvature to vanish at $10^{-1}$ order ($95 \% \, \mbox{CL}$), corresponding to a value of $\alpha$ of order $\gtrsim 500$ (i.e., the largest level set value in Fig. \ref{FLRWtraj}) \cite{Spergel,*Riessetal}.
More recently, precision data from the Planck collaboration, in particular regarding CMB temperature (TT) and polarization (EE) power spectra and CMB lensing, in combination with BAO
constraints, have been taken to imply that $|\Omega_k| \lesssim 5 \times 10^{-3}$ \cite{Adecs}.
Obviously, this corresponds to a huge increase in the lower bound on $\alpha$, i.e., 
\begin{equation}\label{alphaineq}
\alpha \; \gtrsim \; 3,\!000,\!000
\end{equation}
(this more accurate constraint has not been included in Fig. \ref{FLRWtraj}, essentially because the corresponding level sets would blur into the delimiter representing
the flat model)\footnote{Although the curvature constraint - and hence the constraint (\ref{alphaineq}) on $\alpha$ - based on the Planck data is not model independent, there
appear to be good reasons to expect that model-independent constraints at least as strong are a feasible prospect for the near future \cite{Witzemancs}.}.
It should be stressed that the foregoing argument is of general validity, despite the fact that it was set up for the special case of dust, i.e., $w=0$ (the reason for having explicitly considered
this special case is clearly that the argument is then relatively simple and moreover readily visualized). More precisely, given \emph{any} FLRW model that incorporates an arbitrary number of non-interacting
``particle'' species with densities $\rho_i$, satisfying an equation of state of the form $p_i = w_i \rho_i$, with $w_i$ constant and distinct for each species, a complete set of
FLRW constants of the motion can be explicitly constructed \cite{Lake2}.
The crucial point is now that, as long as no species are removed, higher-order FLRW models (i.e., with larger numbers of species) \emph{inherit all constants of the motion}
from corresponding lower-order FLRW models.
Thus, in the physically more accurate case of a FLRW model including both dust and radiation ($w_{\gamma}=1/3$), $\alpha$, as defined by Eq.\ (\ref{FLRWconstant}), is still a constant of the motion
and the fact that, as the Planck data strongly suggest, this constant of the motion is large, i.e., satisfies (\ref{alphaineq}), means that its level surface in the full phase space always remains very close to the
$\Omega_{\mathrm{m}} + \Omega_{\Lambda} + \Omega_{\gamma} = 1$ plane representing the flat FLRW model in this case\footnote{It is worth stressing that the foregoing conclusions are independent of how 
$\Lambda$ is interpreted, i.e., as being associated with geometry or with a $w=-1$ perfect fluid source. In fact, if the latter interpretation is adopted (as has implicitly been done
in the discussion of multi-component FLRW models), the conclusions are invariant under perturbations about $w=-1$ \cite{Lake}.}.\\
A final point worth mentioning is the following. As the preceding discussion shows, the Universe we inhabit happens to be best described by a large-$|\alpha|$ FLRW model.
One could still ask how special such a model is however.
Or, put differently, how generic are FLRW models with $\Omega_{\mathrm{m}} + \Omega_{\Lambda} \simeq 1$ throughout their history?
As was recently shown by Helbig \cite{Helbig,*Helbig2} (recall also the discussion immediately preceding Eq.\ (\ref{Omegatotal}) above), the answer to this latter question is that such models are indeed very generic. In particular, temporally open models that are not always nearly
flat are necessarily elliptic and satisfy $\alpha \approx 1$. In other words, the fine-tuning is reversed: models that are not almost flat throughout their history are necessarily fine-tuned.

\subsection{$\Omega \simeq 1$ as an Anthropic Coincidence}

\noindent A final version of the flatness problem sometimes encountered (although, again, conflations with the two
previously discussed arguments are not uncommon) is based on the observation that spatial flatness appears to be 
an ``anthropic coincidence''. That is, a universe in which $\Omega$ is smaller or larger than unity by significantly more 
than an order of magnitude would (presumably) be \emph{completely} unlike the Universe as actually observed;
in particular, such a hypothetical universe would (presumably) not be consistent with the existence of life in the form we know it (see
e.g.\ Ref. \cite{Carr} and references therein for an argument along these lines).
Similar observations have been made regarding other basic parameters used in physical theory, i.e., Newton's constant, 
$G_{\mbox{\scriptsize N}} = 6.7 \times 10^{-11} \mbox{N} \mbox{m}^{2} \mbox{kg}^{-2}$, the electron charge,
$e = 1.6 \times 10^{-19} \, \mbox{C}$ (or the fine-structure constant, $\alpha := e^2 / 4 \pi \hbar c \simeq 1/137$), 
Fermi's constant, $G_{\mbox{\scriptsize F}} = 1.4 \times 10^{-62} \mbox{J} \mbox{m}^{3}$ and so on. 
The issue as to whether such observations are to be interpreted as constituting an actual problem (and if so, 
what the extent of this problem is) is at the basis of one of the most vigorously debated topics in theoretical physics over the past decades.\\
One could say that the particular (orders of magnitude of) parameter values observed are simply a precondition for our existence -
so that it is not at all surprising that we actually happen to observe these values - and just leave it at that.
This so-called (weak) ``anthropic reasoning'' is perfectly self-consistent, but many physicists who feel that a deeper
explanation of the observed coincidences somehow must exist, have in recent decades, rather controversially, invoked 
strong anthropic arguments to defend a ``multiverse'' concept of physical reality\footnote{This particular way of referring to weak and strong anthropic reasoning,
although in line with some authors (see e.g. Ref. \cite{Penrose6}), does not appear to be standard terminology.
It is however arguably a \emph{sensible} terminology for rather obvious reasons.}.
This is obviously not the place to go into this contentious subject matter in detail , but in relation to the present
discussion two key points should be stressed.\\
First, it might be thought that treating flatness as an ``anthropic coincidence'' in essentially non-dynamical terms is
illegitimate, since $\Omega$, of course, actually \emph{is} a dynamical parameter.
The point here however is not to deny that $\Omega$ is a dynamical parameter, but rather that the fact that it is, 
is irrelevant for anthropic considerations based on the fact that $\Omega \simeq 1$.
After all, the mathematical question of whether a particular value of $\Omega$ at some particular time is in 
some definite sense probable or not is objectively \emph{different} from the metaphysical question of whether one should 
be surprised to observe a particular value for $\Omega$ (or any other relevant parameter), if that value happens
to be a precondition for one's existence.
Moreover, although it is sometimes contended (conflation of different flatness arguments) that \emph{had} $\Omega$ in fact been
``fine-tuned'' at the Planck scale to for instance five decimal places more or less than its presently calculated value,
it would no longer be $\simeq 1$ after $\simeq 10^{17} \mbox{s}$, such a representation of things is obviously incorrect 
in view of the results discussed in subsection \ref{instable}.
Indeed, as those results made clear, FLRW models for which $\Omega \simeq 1$ throughout their entire history are in a 
well defined sense generic and so notwithstanding the fact that the ``anthropic window'' $0.1 \lesssim \Omega \lesssim 10$ may 
appear to be small, values outside this window are in an appropiate sense much less probable (cf. also the earlier example of the 
$60 \%$ window $1 < \Omega_{\mathrm{m}} < 10$ in the cycloidal universe).\\
The second key point about anthropic arguments in connection to the present discussion bears on the wider issue of ``context
of discovery'' versus ``context of justification'' of these arguments, as well as that of the two other pillars of the modern
trinity consisting of string/M-theory, (eternal) inflation and ``multiverse''.
As has become evident from the discussion in this section, in the case of inflation at least part of the former context
is theoretically unsound and this evidently raises questions about the remaining part and about the context of justification.
These issues, together with various pieces of evidence against discovery-context arguments for string/M-theory and
the ``multiverse'', are briefly discussed in the next section.
			
\section{The Issue of Initial Conditions}\label{IC}

\noindent According to Eq.\ (\ref{RWmetric2}), an expanding FLRW model with $\rho + 3P \geq 0$ and $\Lambda \leq 0$ must start off
in a singular state, $a=0$. Indeed, any FLRW model that is expanding at a \emph{constant} rate, becomes past-singular
at a finite prior time $H^{-1}$ and so any FLRW model that is expanding at a \emph{decelerating} rate has to become
past-singular in a time less than $H^{-1}$.
Obviously, this argument is blocked if either $\rho + 3P < 0$ or $\Lambda > 0$.
However, all viable, classical cosmological models known to date satisfy $\rho + 3P > 0$
(which is essentially just another way of saying that these models satisfy the ``strong energy condition'' that the 
stress-energy tensor contracted with any unit timelike vector field is bounded from below by minus one half its trace), 
while even though $\Lambda > 0$ for the observable Universe in its present state, this does not in fact prohibit reversing the trend of 
contraction in the past time direction.
In other words therefore, all cosmologically relevant classical FLRW models start with a ``Big Bang'' a finite amount of time ago\footnote{In fact, 
the foregoing line of argument, based on Eq.\ (\ref{RWmetric2}) is a bit deceptive, as it might superficially appear that a 
past-singularity could be avoided in any expanding FLRW model, provided $\Lambda$ is \emph{large} enough, whereas the 
discussion in section \ref{isotropy} should make clear that exactly the opposite is the case. Indeed,
any dynamical trajectory in the upper-right quadrant of the $a\Lambda$-plane for which $\dot{a}$ and $\ddot{a}$ are
both positive at some instant, in the reversed time direction (i.e., as obtained by following the appropriate 
constant-$\Lambda$ line to the left) either hits a singularity at $a=0$  or a ``bounce'' at some finite $a$-value,
where $\dot{a}$ changes sign, but the latter can occur only for $k=+1$ and $0 < \Lambda < \Lambda_{\mathrm{E}}$.
Although it may not seem obvious that the inevitability of a past-singularity persists in the case of an empirically
adequate multi-component perfect fluid FLRW model, it in fact does and moreover so even without the need to include
any prior assumptions about the value of $\Lambda$ \cite{EhlRin}.\label{BBinevitable}}.
As is well known, according to one of the celebrated Hawking-Penrose singularity theorems \cite{HawkingPenrose}, this conclusion 
extends to \emph{all} general relativistic models that are in some appropriate sense ``physically reasonable''.
In other words, given certain physical conditions that, as all evidence suggests, pertain to the observable Universe
(at the classical level at least), the high-symmetry FLRW-context is completely representative, in that general relativity
predicts the existence of an initial spacetime singularity - and thereby, unavoidably as it seems, its own breakdown\footnote{More
precisely, the archetypal singularity theorem states that any general relativistic spacetime that satisfies (i) some 
appropriate causality  condition (for instance, the absence of closed timelike curves), (ii) some appropriate energy 
condition (for instance, the strong energy condition) and (iii) some appropriate focussing condition (for
instance, the existence of a so-called ``trapped surface''), is necessarily \emph{inextendible}. That is, such a spacetime necessarily
contains at least one incomplete geodesic (i.e., a geodesic with only a finite parameter range in one direction, but
inextendable in that direction) and is not isometric to a proper subset of another spacetime.
It is important to note that, whereas it is very well conceivable that some of the above conditions may not strictly
be satisfied at all spacetime points, it does not appear that - within a classical gravitational context - one could thereby
avoid spacetime singularities in any realistic sense.
For instance, violations of the (strong) energy condition can occur in the very early Universe because of the effects of 
quantum fields and/or inflationary theory, but if the condition holds in a spacetime average sense, singularities are 
still expected to occur, in general \cite{Joshi}.
That the conditions typically assumed by the singularity theorems are merely sufficient conditions can also be understood
from the interpretation of the cosmological constant as a negative-pressure perfect fluid. The net effect of that
interpretation is to contribute towards violating the strong energy condition (for positive $\Lambda$), but this has
in itself little bearing on the issue of the occurrence of spacetime singularities.
For instance, as is easily verified from the discussion in section \ref{isotropy}, for $\Lambda$ sufficiently large, 
all non-empty isotropic models on this view violate the strong energy condition most of the time, 
but are nevertheless past-singular (recall also note \ref{BBinevitable}).}.
This Big Bang account of the origin of the Universe moreover receives strong further support from various pieces
of observational evidence, such as cosmic abundances of the light elements, i.e., (isotopes of) H, He and Li, pointing to an era during which
nucleosynthesis occurred, about 13.8 billion years ago, and the relic photon distribution from the ``primordial
fireball'', i.e., the CMB, which formed at about 380,000 years after nucleosynthesis.
In combination, these various pieces of theoretical and empirical evidence form a practically conclusive, coherent overall 
picture, according to which the Universe contracts further and further in the past time direction, with densities and temperatures reaching higher
and higher levels, all the way up to \emph{at least} the scale at which, according to the standard model of particle physics,
electroweak symmetry was ``broken'', about a million millionth of the time it took for nucleosynthesis to begin and when temperatures reached values of order $10^{15} \, \mbox{K}$.
\enlargethispage{1cm}

\subsection{The Extra-Ordinary Special Nature of the Big Bang}

\noindent It is generally assumed that a future theory of quantum gravity will somehow resolve the (initial and final) singularities 
implied by general relativity, but whether that will turn out to be the case and, if indeed so, what it would entail
for the scale at which the above classical picture exactly breaks down, is not too important for the present discussion.
In fact, other authors have given different estimates of scales down to which the Big Bang account should (at least) be 
trustworthy (for instance, the scale at which, according to some models, the Universe underwent
a ``GUT symmetry breaking transition'' and which supposedly took place at some $10^{-33} \, \mbox{s}$ \cite{Ellis2}) 
and the claim that the account should be reliable down to at least the 
electroweak scale at $\tau_{\mbox{\tiny EW}} \simeq 10^{-12} \, \mbox{s}$, seems rather conservative. 
Moreover, from a cosmological perspective, the electroweak scale is rather special and may in fact hold some important 
clues about the state of the Universe at ``earlier times'' \cite{Holman1}.\\
The key point about the foregoing discussion however is that, to whatever microscopic scale the Big Bang picture implied
by general relativity is trusted, the ``initial condition'' of the Universe corresponding to that scale was
\emph{extra-ordinarily special}.
Indeed, as already mentioned before, deviations from exact uniformity at ``decoupling'', $\tau_{\mbox{\tiny d}} \simeq 380,\!000 \; \mbox{years} \simeq 10^{13} \, \mbox{s}$,
when electrons and nuclei combined into neutral atoms and the Universe became transparent to electromagnetic radiation, amounted to only a few parts in $10^5$, whereas general
reasoning based on the second law of thermodynamics implies that, at earlier times, deviations from uniformity were
even smaller (the paradox that usually in thermodynamics, systems tend to become more uniform towards the \emph{future} - as
with the familiar gas in a box - arises precisely because for such systems the effects of gravitational clumping are utterly negligible).
To get a sense of just \emph{how} extra-ordinarily special an event the Big Bang was, Penrose \cite{Penrose1,*Penrose5} has taken the total entropy
of matter estimated to have collapsed into black holes at present as an upper bound for the entropy of the initial state
and by comparing this entropy to that of a black hole of a mass of the order of the (observable) Universe, derived an upper bound for the
fractional phase space volume representing the Big Bang, which is of order
\begin{equation}\label{BBentropy}
\mbox{one part in} \; 10^{10^{123}}
\end{equation}
On this view, there is thus indeed a problem of fantastic ``improbability'' corresponding to the Universe's initial state
(and it is clear from the above formulation that the degree of unlikeliness is completely immune with respect to whether this
initial state is envisaged to take effect at the Planck scale, the electroweak scale, the ``decoupling scale'', or,
in fact, \emph{any} scale corresponding to some prior moment in the Universe's history for that matter), but in 
sharp contrast to the discussion of section \ref{FP}, this problem is primarily concerned with the apparent ``improbability''
of a FLRW initial state \emph{in general}. It is thus \emph{different} from the alleged flatness problem, since, as seen
explicitly in section \ref{FP}, \emph{any} physically relevant FLRW initial state necessarily will have $|\Omega - 1|$ extremely close to $0$.\\
In fact, it is this apparent unlikeliness of an isotropic initial state that underlies a second major putative issue 
generally associated with FLRW cosmologies, which, as already alluded to before, is the general appearance of
\emph{particle horizons} in such cosmologies. The essential point of concern is that because of the spacelike nature
of the initial singularity, most regions on the celestial sphere of vision of a given isotropic observer at some particular time,
correspond to portions of the decoupling (i.e., ``last scattering'') surface, $\Sigma_{\tau_{\mbox{\tiny d}}}$, with causally disconnected histories
and it is then argued to be problematic that, when comparison is made to the real Universe (i.e., through the CMB), 
physical conditions in all these portions of $\Sigma_{\tau_{\mbox{\tiny d}}}$ should nevertheless have been so extremely similar.
The difficulty with this argument however is that, again, there is really no conceptual puzzle here within a strict
FLRW context, since the uniformity is simply present from the very beginning by assumption (i.e., $\rho$ is \emph{constant}
on \emph{each} hypersurface of homogeneity, $\Sigma_{\tau}$).
Of course, this is only a zeroth order approximation (except, arguably, at primordial times), but
it is clear that if one were to (reasonably) argue that particle horizons should still be present for cosmological
perturbations sufficiently close to exact FLRW symmetry, there is a priori no reason to suppose that for such
perturbations there \emph{would} be a conceptual difficulty brought about by the empty intersection of past domains of
dependence of typical portions of $\Sigma_{\tau_{\mbox{\tiny d}}}$ in an observer's past lightcone, since deviations in the matter
density, $\delta \rho$, from exact uniformity would \emph{also} have to be sufficiently small, again by assumption (i.e.,
$\rho$ would have to be ``sufficiently close'' to being constant on each $\Sigma_{\tau}$).\\
From a general perspective, it is a prima facie open question whether particle horizons still arise within an 
anisotropic  (e.g.\ merely spatially homogeneous) framework. In fact, it was originally 
argued by Misner \cite{Misner} that particle horizons are \emph{absent} in homogeneous cosmologies of Bianchi type IX  
and that, as a result, such cosmologies could start off in a ``sufficiently generic'' state, which would then subsequently 
``thermalize'' into a near-isotropic state before decoupling.
Although this particular idea of a ``Mixmaster model'' turned out not to work, what the foregoing remarks amply demonstrate
is that there are in fact several \emph{different} issues at play in what is usually referred to as the ``horizon problem''
of the standard FLRW cosmologies (which, as noted above however, is strictly a terminological inconsistency).
There is a genuine conceptual/physical difficulty in this regard only if it is \emph{assumed} that in fact
\begin{itemize}\label{HPassumptions}
\item[(i)] the initial state, say at the Planck time, was somehow ``sufficiently \emph{non}-uniform'', with ``sufficiently
strong'' density fluctuations, $\delta \rho$, and
\item[(ii)] no ``ordinary'' dissipative processes could have effectively isotropized this sufficiently non-uniform initial state by
the time the CMB formed, i.e., when matter and radiation decoupled, \emph{without} postulating new forms of matter (fundamental
or effective), either because of the existence of particle horizons or because of some other reason.
\end{itemize}
Regarding the first assumption, as seen in section \ref{isotropy}, all current cosmological data point to an essentially isotropic
large-scale Universe between decoupling and the present epoch - i.e., for effectively the Universe's \emph{entire} history
(the time it took for radiation and matter to ``decouple'' obviously being insignificant relative to the Universe's present age).
So, based on just that elementary observational fact, it would not seem an enormously outlandish extrapolation to
assume that the isotropy extends all the way back to the Big Bang. In fact, according to general thermodynamic reasoning,
this is exactly what is to be expected: tracing the Universe's evolution backwards from decoupling time, when temperature
and density anisotropies were of order $10^{-5}$, elementary considerations based on the (generalized) second law of
thermodynamics indicate that the Universe's decreasing entropy can for the most part be attributed to further suppression
of gravitational degrees of freedom, since matter and radiation were essentially always in thermal equilibrium.\\
On this view, the Universe thus becomes ever more isotropic in the past time direction and as the initial state is approached,
the FLRW approximation becomes essentially exact.
Obviously, the key question then becomes \emph{why} the very early Universe was so extra-ordinarily
special, i.e., as conservatively expressed by the fantastically small phase space volume (\ref{BBentropy}).
Or, put slightly differently, \emph{what} physical processes constrained entropy to be so extremely low at the Big Bang?
This question certainly seems very far from being answered today. Moreover, it is also certainly not intended
to suggest here that, while this general type of explanatory account of the orgins of the second law of thermodynamics -
that is, by tracing these origins to boundary conditions - has become standard,
such an account is free of technical or conceptual difficulties \cite{Penrose1,*Penrose5}, \cite{Feynman}, \cite{Wald3}.
However, there are good reasons to think that a future theory of quantum gravity would be exactly what is needed to
address these types of issues and that, moreover, from a methodological perspective, the search for such a theory should
be fundamentally guided by seeking an explanation of the ultra-low entropy Big Bang in the form of a \emph{lawlike} initial
condition \cite{Holman2}.\\
In spite of these points however, it is fair to say that most physicists at 
the present time, through their endorsement of the ``cosmological concordance model'', would - either implicitly or 
explicitly - argue in exactly the opposite direction.
That is, rather than accepting what the primary evidence strongly seems to suggest (and thereby being placed in the 
position to account for the thermodynamical arrow of time - one of the most basic facts about physical experience), they
would subscribe to the view that the initial state was \emph{not} of the uniform type, because of its extreme ``unlikeliness'',
precisely as expressed by (\ref{BBentropy}) (or, in mathematically more sophisticated jargon, by the statement that 
FLRW-type singularities have ``zero measure'').
In fact, a folk theorem in general relativity states that generic singularities belong to a particular class, which, 
following Penrose, will here be referred to as (the class of) \emph{BKLM singularities} \cite{BKL}, \cite{Misner}.
According to this view, a \emph{generic} (spacelike) initial singularity would be highly non-uniform - essentially
``chaotic'' - and exhibit an incredibly complicated, but locally homogeneous, ``oscillatory'' behaviour at ``infinitesimal
distances away'' from it\footnote{More accurately, according to the BKLM picture, apart from zero measure counter-examples,
all spacelike initial singularities in general relativity are ``vacuum dominated'' (i.e., ``matter does not matter'' - that 
is, with the apparent exception of possible scalar matter), ``local'' and ``oscillatory'' (in the sense of being locally 
homogeneous, i.e., Bianchi type, and approached through an infinite sequence of alternating ``Kasner type epochs'', as 
first described in a more specific context within Misner's Mixmaster model \cite{UEWE}).
It should be stressed however that this picture at the present time very much amounts to an unproven \emph{conjecture}
and that specific model studies have provided evidence both for and against it (for a recent discussion of the status of 
the conjecture, see Ref. \cite{Berger}).
One of the prima facie obstacles in obtaining a proof of the conjecture would appear to be making quantitative the notion 
of ``genericity'' (i.e., through defining a meaningful measure on the space of all singular cosmological models).
However, as will become clear in the main text, even if the conjecture were to hold in the sense that generic
spacetime singularities would essentially be of BKLM type, it is rather doubtful that this would establish anything
about the \emph{actual} Big Bang.
Indeed, within a purely classical context, there is no reason to treat initial and final spacetime singularities within
the mathematical formalism of general relativity differently, but that does not mean that there are no \emph{physical} 
motivations to do so. In fact, the thermodynamic arrow of time very strongly amounts to precisely such a motivation.}.
This would thus mean that the first of the two above assumptions necessarily (although usually adopted tacitly) underlying
the received view with regards to the existence of an actual horizon problem in cosmology is amply satisfied.
Regarding the second assumption, it was established already some time ago that Bianchi models in general do not isotropize
at asymptotically late times (in fact, the homogeneous models that do isotropize were found to have zero probability) \cite{ColHaw}.
However, it turns out that for instance dust- or radiation-sourced models of Bianchi type $\mbox{VII}_h$ with non-zero
probability can isotropize at \emph{intermediate} times, in a manner that appears to be fully consistent with observations
of the CMB \cite{WCEH}.
Nevertheless, on accepting the seemingly conventional wisdom based on both assumptions (i), (ii), it is clear that
there is an actual physical problem and it is this problem which the $\Lambda \mbox{CDM}$ cosmological concordance model
intends to resolve by postulating the existence of a new quantum particle without spin, $\phi$, the so-called ``inflaton''\footnote{Although one could
continue to speak of a ``horizon problem'' if conditions (i), (ii) are satisfied, it would seem more appropriate to refer
to the underlying effective isotropy at ``late'' times as being problematic; cf. section \ref{epilogue}.
It is also worth noting that, given that quantum (effective) fields had to be dominant in the very early Universe, it does not at all 
seem clear that isotropization could \emph{not} have occurred sufficiently rapidly through \emph{non-ordinary} dissipative processes involving 
the extreme spacelike entanglement of these fields \cite{Wald4}.}.

\subsection{Inflationary Cosmology}

\begin{quote}
Some physicists (a century ago) suggested that all that has happened is that the world, this system that has been going
on and going on, fluctuated. It fluctuated, and now we are watching the fluctuation undo itself again $[\cdots]$
I believe this theory to be incorrect. I think it is a ridiculous theory for the following reason.\\
R. P. Feynman \cite{Feynman}
\end{quote}

\noindent Now, it should be stressed that the inflationary models mentioned in subsection \ref{finetune}, that were originally
introduced as solutions to the (alleged) flatness and horizon problems of the standard FLRW cosmologies, are \emph{not}
based on the BKLM picture of the Big Bang, but effectively \emph{assumed} isotropy to be present from the very beginning.
In fact, in these models the inflationary epoch was supposed to have been preceded by a ``GUT symmetry breaking'' transition
that allegedly took place some $10^{-33}$ seconds after the Big Bang and during which certain types of magnetic monopoles 
should have been copiously produced \cite{Preskill}.
Since no such monopoles had ever been observed, whereas the ``Grand Unified Theories'' (i.e., GUT's) that predicted them
were generally regarded as having high theoretical standing, this presented a serious problem.
According to the original inflationary models, this problem would be resolved if the very early Universe, immediately
after the GUT transition, would have gone through an extremely brief period of \emph{exponential expansion} - thereby 
effectively behaving as an empty FLRW model with positive cosmological constant (i.e., a de Sitter spacetime) - during 
which the scale factor of the FLRW metric ``blew up'' by an absolutely stupendous factor (typically in the order of a 
googol, i.e., $10^{100}$ or so)\footnote{Whether the expansion factor is actually a googol, the square root of a googol, or a googol squared, etc.
is not important for present considerations and in fact it does not appear that inflationary models themselves
are currently able to make a firm prediction in this regard.
It should also be noted here that the original inflationary models \cite{Guth1,*Linde3,*Albstein} mentioned earlier
were in fact somewhat different from each other. In Guth's 1981 model (now usually referred to as ``old inflation''), the exponential
expansion was envisaged to take place with the inflaton sitting at the top of the Mexican hat potential, so to say ($\dot{\phi}$
constant), whereas in the Linde-Albrecht-Steinhardt 1982 models (now usually referred to as ``slow roll (or new) inflation''),
the expansion occurred with the inflaton rolling down from the top of the sombrero potential ($\dot{\phi}$ time-varying).
Note however that, in view of the fact that these initial models were formulated in an (effectively) isotropic context to
start with, it is rather doubtful (given in particular the dynamical behaviour of $\Omega$ and the horizon issue, as 
discussed at length earlier), that they addressed any internal problem of cosmology that existed in the first place.}, 
concomitantly diluting any monopole densities present to acceptably small values along the way.
Following the proton-decay non-observations in the early 1980s, which ruled out a nontrivial number of popular, simple GUT models
(such as the non-supersymmetric $\mbox{SU}(5)$ model that reportedly \emph{should} have been correct, based purely on its 
apparent aesthetic qualities), as well as the realization that inflationary models appeared to be rather special (as later 
studies amply confirmed \cite{SteinTurn,*RoEl,*RayMo,*MaEl,*Penrose2}), it however soon became clear that a more generic picture of the conditions prior to inflation was required.\\
This led in particular to the notion of ``chaotic inflation'' \cite{Linde4}, which was essentially based (at least in spirit)
on some sort of BKLM picture of the initial state\footnote{Although no explicit mention is made of the BKLM conjecture, Ref. \cite{Linde4}
does speak of a standard view according to which the Universe before the Planck era is ``in some chaotic quantum state''
and at least some of the cited advocates of this view do approvingly refer to (part of) the conjecture (see e.g.\ Ref. \cite{Rees}).
However, in the actual model of Ref. \cite{Linde4} chaos does \emph{not} in fact enter spatial geometry in any way.
The model is chaotic only in the sense that arbitrary constant initial values of the $\phi$ field (within some
specified range) are allowed for different spatial regions (which are themselves isotropic and of typical dimensions much larger than the 
Planck length) at whatever time inflation is supposed to start.
Nevertheless, since the consensus view appears to be that the BKLM conjecture holds, it seems that a genuinely chaotic inflationary
model should also refer to chaotic initial conditions in spatial geometry (for work in this direction, see e.g.\ Ref. \cite{HuOC,*FuRoMa}),
\emph{especially} within a framework that is at its core time-reversal invariant, as is usually taken to be the case \cite{Albrecht}.
It should also be stressed that the picture of a ``chaotic ensemble'' of Planck order sized patches sketched in the main text 
is intended for heuristic purposes only, since a clearly defined framework for these inflationary patches - let alone a
mathematically rigorous treatment - is currently unavailable.}.
From a modern vantage point and skipping some technical details, the essential picture underlying chaotic inflation can be described as follows.
In accordance with the BKLM proposal, the state of the Universe at Planckian times $\simeq 10^{-44} \, \mbox{s}$,
where nontrivial quantum gravity effects are expected to play a dominant role, is pictured as a highly complex, foamlike ``chaos'' of
Planck scale sized patches.
The overwhelming majority of these patches would never inflate and would evolve into spacetime structures that are not even
remotely similar to the observable Universe. In some regions however, conditions would be \emph{just} right
for inflation to occur and consequently, these regions would effectively isotropize on exponentially rapid timescales\footnote{It
has been demonstrated that a generic, initially expanding Bianchi model with positive cosmological constant inflates into an effective de Sitter 
spacetime within an exponentially short timeframe \cite{Wald5}.
This ``cosmic no-hair'' theorem however crucially depends on the assumption that the matter stress-energy tensor satisfies
the strong energy condition, which seems physically unreasonable in the very early Universe.
In particular, it has been pointed out that the de Sitter solution is unstable if the condition is violated \cite{Barrow}.}.
It is certainly a \emph{possibility} therefore, within this kind of picture, that the high isotropy of the large-scale
Universe in its present state came about through an extremely brief period of inflationary expansion of an appropriately
conditioned initial patch.
The problem with this argument however, is that it is \emph{also} a possibility, within this kind of picture, that
the present Universe just evolved from an appropriately conditioned initial patch within the spacetime quantum foam
\emph{without} inflation\footnote{Such an initial patch could have been either already isotropic or anisotropic (as in the 
scenario of Ref. \cite{WCEH} and more in accordance with the BKLM picture).}.
In fact, if the Universe's spatial geometry is infinite, both scenarios are bound to occur, whereas arguments
that purport to establish that the inflationary scenario would somehow be \emph{significantly} more probable, 
inevitably seem to run into the pitfall of introducing a host of further, unverifiable assumptions, in such a way that the input always seems to be far
greater than the output\footnote{It might be thought that cosmological precision tests should in practice be able to shed light on the question as to whether
the early Universe actually went through an inflationary epoch, for instance by determining the parameter values of the
inflaton potential. The problem is however that virtually \emph{any} set of relevant observational data can be matched 
with an appropriate inflationary model (a situation that actually pertains to both the dynamical evolution of the scale 
factor and to the spectrum of density perturbations; see e.g.\ Refs. \cite{Ellis2}, \cite{Padmanabhan}).\label{inflationspectrum}}.\\
It appears that the reason why these fundamental difficulties have nevertheless not deterred many theorists from
actively promoting the inflationary picture as part of established physics (viz. the cosmological concordance model)
is inextricably related, at least in part, to what is usually advertized as another key future of the chaotic inflationary
discourse, namely the idea that such a discourse inevitably leads to the notion of (future) \emph{eternal inflation}.
According to this notion, once inflation has started somewhere (i.e., in some fantastically atypical patch), it will
necessarily keep on going forever, producing literally infinitely many ``pocket universes'' (or inflationary ``bubbles'')
in the process \cite{Guth2}.
Many theorists apparently interpret such a cartoon-like picture as some kind of existence proof of a physical multiverse
of the kind that has also been argued for in string/M-theory, particularly in the past decade and a half (i.e., based on the
infamous ``string theory landscape'').\\
Returning to the issue raised near the end of section \ref{FP}, although the previous discussion shows that the case
for inflation based on its capacity to resolve the issues it was originally designed for, is virtually non-existent,
it is of course true that inflationary models have been spectacularly successful in accounting for slight \emph{deviations}
from exact uniformity in the CMB spectrum.
In particular, it is widely taken to be the case that inflation generically predicts that the spectrum of primordial
density perturbations is (almost) \emph{scale-free} and \emph{Gaussian}.
The trouble with presenting this feature as providing conclusive evidence for inflation \emph{a posteriori} however is that
exactly the same density perturbation spectrum can be obtained from non-inflationary models \cite{HollandsWald1} (recall
also note \ref{inflationspectrum}); in fact, the specific prediction of a scale-invariant spectrum of density perturbations
was made long before inflationary models even existed. As far as predictions are concerned that are believed to be true
signatures of inflationary models (such as so-called \emph{spectral indices} that for instance characterize minute \emph{deviations}
from scale invariance), opinions diverge as to whether such predictions are in agreement with the most recent observational data \cite{ISL,*ISL2}. 
With regards to the justification of the strong anthropic principle, it has become clear in recent years that some fine-tuning arguments - in particular, those
involving the cosmological constant and the Higgs mass - which were at times interpreted as \emph{independently} suggesting, 
or even providing strong evidence for, the validity of this principle, are actually far less conclusive than previously
taken to be the case \cite{HollandsWald2}, \cite{Holman1}.\\
Finally, it is often not realized that the ultra-low entropy of the Big Bang, as expressed by (\ref{BBentropy}), constitutes
a powerful argument \emph{against} the idea that the early Universe actually inflated.
The point is that, as seen in detail earlier, the modern notion of ``chaotic inflation'' is based on the premise, either
in theory or in practice, that inflation started in an essentially random state.
Thus, rather than being constrained to the extra-ordinary small fractional phase space volume as expressed by (\ref{BBentropy}),
the Big Bang, in this scenario, being riddled with high-entropy \emph{white holes} (i.e., time-reverses of black holes),
is taken to be completely unconstrained. In other words, on this view, our entire 13.8 billion year old Universe in all its
various manifestations, is nothing but a staggering ``fluctuation'' from an ``eternal'' primordial thermal equilibrium state.
The problem with this view, again, is that while it perhaps may not seem utterly implausible a priori on somewhat 
superficial grounds, it doesn't in fact explain \emph{anything} about the \emph{actual} Universe we inhabit.
Taking the time-honoured case of standard thermodynamics, some intuitive meaning can be attached to a 
thought experiment in which a fictitious (and extremely patient) experimentalist waits around to observe a gas at thermal equilibrium 
inside, say, a box of $1 \, \mbox{m}^3$, spontaneously fluctuate into a region a thousandth times the box's volume, every
$10^{10^{26}}$ years or so (i.e., within a Poincar\'e recurrence time corresponding to the gas molecules 
originally confined to such a volume). 
The outcome of such a thought experiment seems a perfectly feasible explanation of the fact that 
spontaneous fluctuations of this type are in fact never observed in our actual Universe.
But for the thought experiment, it would also be understood that on those exceedingly rare occasions where the entire 
gas spontaneously fluctuated into a volume a thousandth times its size (or any small volume for that matter),
it would not \emph{stay} in such a low-entropy state long.
In sharp contrast, a primordial thermal equilibrium state spontaneously fluctuating into an ultra-low entropy Universe,
which is then to be governed by apparently thermodynamic, second law-like behaviour for at least 13.8
billion years, giving rise to (conscious) life along the way, while all the time effectively retaining its
overall ultra-low entropy, is an utterly fantastical idea which is completely at odds with how our Universe is actually observed to behave.\\
As Penrose has long emphasized and contrary to what is assumed in the chaotic inflation picture (either 
in theory or in practice), the thermodynamic arrow of time provides a very powerful reason to think that the Big Bang 
was \emph{not} in a state of thermal equilibrium.\enlargethispage{1cm}
This thus means that the BKLM picture of the Big Bang as essentially a white hole riddled, ultra-high entropy state
is false for \emph{physical} reasons and that a future theory of quantum gravity should come equipped with a fundamental
arrow of time \cite{Penrose1,*Penrose5}, \cite{Wald2}, \cite{Wald3}, \cite{Holman2}.
It also means that quantum gravity should somehow provide a lawlike constraint on the initial singularity to be (effectively)
of FLRW type, with vanishing Weyl curvature.

\section{Epilogue}\label{epilogue}

\noindent To return to the question posed in the title, given a global proper time ordering, it is not surprising
that galactic observers should find spatial sections (strictly not including themselves or other galactic 
worldlines, so as to avoid final singularities) to be locally Euclidean, i.e., locally flat, this being an in-built premise of general relativity.
A problem with taking the (normalized) density parameter, $\Omega$, in a FLRW model, as a measure of deviation from
Euclideanity on more global scales, consists in the fact that \emph{all} singular FLRW models start off at the critical
value, $\Omega = 1$, that characterizes a Euclidean model ($k=0$) for all times, whereas the non-Euclidean models ($k=-1$, $k=+1$)
also start off with infinite (scalar) curvature.
In other words, at very early times the non-Euclidean models are at the same time arbitrarily close to being flat, 
as judged from the $\Omega$-parameter, and arbitrarily far from being flat, as judged from the scalar
curvature. This shows that the empirical fact that $\Omega \simeq 1$ in the present epoch, does not a priori imply
anything about the value of $k$ (which, after all, is a discrete parameter) - and hence for the question whether
the large-scale spatial geometry is Euclidean or not (at any rate, the circumstance that all FLRW spacetimes, being
conformally flat, have the same causal structure as Minkowski spacetime, also indicates that the issue of
deciding between flatness and curvature for such spacetimes is a bit more subtle than what mere arguments based on the
value of $\Omega$ would appear to suggest).
Nevertheless, if the standard practice of interpreting $\Omega$ as a measure of Euclideanity, i.e., flatness, is followed, 
it was shown in detail in section \ref{FP} that the current proximity of $\Omega$ to $1$ does not pose an \emph{independent}
flatness problem for cosmology (in particular, all singular FLRW models
are inherently ``fine-tuned'', whereas the circumstance that $\Omega$ appears to be somewhere in the middle of the
anthropic window is no more or less puzzling than similar circumstances for other physical parameters).\\
In addition, it was seen in detail in section \ref{IC} that there is an actual ``isotropy problem'' (of a very definite
nature) for the present epoch, and even more so for the time when matter and radiation decoupled, \emph{only} if it is
\emph{assumed} that the initial conditions of the Universe were sufficiently \emph{non}-uniform (albeit, due to the
anomalous thermodynamic behaviour of gravity, essentially at thermal equilibrium) \emph{and} if no known physical
processes could have effectively isotropized (mainly the gravitational degrees of freedom in) these initial conditions
by decoupling time (for instance, because of the existence of particle horizons; cf. conditions (i), (ii) on 
p. \pageref{HPassumptions}).
But if both these conditions are satisfied and unknown physics is not called upon, this just means that the Universe
could not have been nearly isotropic at decoupling time (in sharp contrast with actual observations, of course).
In particular, it would then be meaningless to argue that there would be an independent problem caused by the fact that
FLRW models would ``retrodict'' $\Omega$ to be fine-tuned to $1$ to six or so decimal places at decoupling, since such
a retrodiction would be based on something already known to be false in this context in the first place, i.e., the premise
that the Universe at decoupling was nearly isotropic. On the other hand, as the discussion in section \ref{FP}
clearly shows, if there is no ``isotropy problem'' in the specific above sense, there can also be no problem caused by
any ``fine-tuning'' of $\Omega$.\\
In the current $\Lambda \mbox{CDM}$ cosmological concordance model, both conditions underlying the isotropy problem
in the above sense are assumed (often tacitly) and new physical processes, in the specific form of scalar field generated
inflationary expansion, are called upon to resolve the problem thus generated.
As argued at length in section \ref{IC} however, there are many reasons why this inflationary element of the concordance
model is unsatisfactory.
In addition, it was argued that there \emph{are} in fact powerful reasons, based on the second law of thermodynamics, to think
that the non-uniformity assumption (i), that leads to a particular problem in the first place, is false for
the actual Universe. Moreover, it was seen that assumption (ii) can be violated (e.g., intermediately isotropizing
Bianchi type $\mbox{VII}_h$ models seem to be consistent with observations).
These remarks however do \emph{not} imply that the isotropy of the current epoch is problematic only if in fact one
assumes it to be so in the specific sense above. In fact, there is broad consensus amongst contemporary researchers that the key problematic issue
underlying the FLRW approximation is its very effectiveness.
Indeed, why \emph{should} the large-scale Universe at present, and even more so at decoupling, be so highly isotropic,
given that the (time-symmetric) mathematical formalism of general relativity would seem to suggest that it must have been
completely different in its earliest stages? Or if, as argued for here, this is not the case, \emph{what} physical
law caused the initial state to have such ultra-low entropy (\ref{BBentropy})?



\newpage
\section*{Acknowledgements}

\noindent This publication was made possible through the support of a grant from the John Templeton Foundation.
The opinions expressed in this publication are those of the author and do not necessarily reflect the
views of the John Templeton Foundation.
I thank the organizers of the \emph{Fourth International Conference on the Nature and Ontology of Spacetime}, in
Varna, Bulgaria, for providing an opportunity for me to present (most of) this work.
Financial support from Stichting FOM (Foundation for Fundamental Research on Matter) in the Netherlands to attend the 
Conference is also gratefully acknowledged.
Finally, I thank Andy Albrecht, Feraz Azhar, George Ellis, David Garfinkle and especially Phillip Helbig and Chris Smeenk for comments 
and/or discussions pertaining to the contents of this manuscript.

\vspace{-2ex}
\bibliographystyle{mltplcit}
\bibliography{flatness}

\end{document}